\RequirePackage{fix-cm}
\documentclass[twocolumn]{svjour3}          %
\usepackage{natbib}
\usepackage[utf8]{inputenc}
\usepackage[T1]{fontenc}
\smartqed  %
\usepackage{graphicx}
\usepackage{mathptmx}      %
\usepackage{latexsym}
\usepackage{amsmath}
\usepackage{nicefrac}
\usepackage{bm}
\usepackage{tikz}
\usepackage{import}
\usepackage{pgfplots}
\usepgfplotslibrary{fillbetween}
\usepgfplotslibrary{groupplots}
\usepgfplotslibrary{colormaps}
\usepackage[scientific-notation=true]{siunitx}
\usepackage{subcaption}
\usepackage{packages/lnmtikz/lnmtikz}
\usetikzlibrary{positioning}
\usetikzlibrary{matrix}
\usetikzlibrary{backgrounds}
\usepackage[misc]{ifsym}
\newcommand*{\affaddr}[1]{#1} 
\newcommand*{\affmark}[1][*]{\textsuperscript{#1}}

\newcommand{\reva}[1]{#1}
\newcommand{\revb}[1]{#1}
\newcommand{\revc}[1]{#1}

\pgfplotsset{
    colormap={paraviewcolormap}{
        rgb255=(59,76,192)
        rgb255=(221,221,221)
        rgb255=(180,4,38)
    },
}
\usepackage{xcolor}
\definecolor{dia3x1}{HTML}{ffa600}
\definecolor{dia3x2}{HTML}{bc5090}
\definecolor{dia3x3}{HTML}{003f5c}

\definecolor{dia4x1}{HTML}{003f5c}
\definecolor{dia4x2}{HTML}{7a5195}
\definecolor{dia4x3}{HTML}{ef5675}
\definecolor{dia4x4}{HTML}{ffa600}

\definecolor{dia5x1}{HTML}{003f5c}
\definecolor{dia5x2}{HTML}{58508d}
\definecolor{dia5x3}{HTML}{bc5090}
\definecolor{dia5x4}{HTML}{ff6361}
\definecolor{dia5x5}{HTML}{ffa600}

\definecolor{diav1x1}{HTML}{003f5c}
\definecolor{diav1x2}{HTML}{8699aa}
\definecolor{diav2x1}{HTML}{ffa600}
\definecolor{diav2x2}{HTML}{ffd291}
\newcommand{\vct}[1]{\boldsymbol{#1}}
\newcommand{\mat}[1]{\boldsymbol{#1}}
\newcommand{\pd}{\partial}
\newcommand{\dd}{\mathrm{d}}
\newcommand{\T}{\text{T}}
\newcommand{\tr}{\text{tr}}

\makeatletter
\newcommand{\gettikzxy}[3]{%
\tikz@scan@one@point\pgfutil@firstofone#1\relax
\edef#2{\the\pgf@x}%
\edef#3{\the\pgf@y}%
}
\makeatother

\pgfplotsset{
    legend style={
        draw=none,
        fill=none,
        font=\small
    },
    legend cell align={left},
    colorbar style={%
        width=0.2cm,%
    }
}
\begin{document}

\title{A homogenized constrained mixture model of cardiac growth and remodeling: Analyzing mechanobiological stability and reversal
}

\author{Amadeus M. Gebauer\protect\affmark[1]         \and
    Martin R. Pfaller\protect\affmark[2] \and Fabian A. Braeu\protect\affmark[3,4] \and Christian J. Cyron\protect\affmark[5,6] \and Wolfgang A. Wall\protect\affmark[1] %
}

\authorrunning{Amadeus M. Gebauer et al.} %

\institute{\Letter $\;\;$ Amadeus M. Gebauer\\
    \email{amadeus.gebauer@tum.de}\\
    Tel.: +49 89 289 15255 \\\\
    \affaddr{\affmark[1] Institute for Computational Mechanics, Technical University of Munich, 85748 Garching b. München, Germany}\\\\
    \affaddr{\affmark[2] Pediatric Cardiology, \revc{Stanford Maternal \& Child Health Research Institute}, and Institute for Computational and Mathematical Engineering, Stanford University, Stanford, USA}\\\\
    \affaddr{\affmark[3] Singapore-MIT Alliance for Research \& Technology Center, Singapore}\\\\
    \affaddr{\affmark[4] \revc{Singapore Eye Research Institute (SERI), Singapore}}\\\\
    \affaddr{\affmark[5] Institute of Continuum and Materials Mechanics, Hamburg University of Technology, 21073 Hamburg, Germany}\\\\
    \affaddr{\affmark[6] Institute of Material Systems Modeling, Helmholtz-Zentrum Hereon, 21502 Geesthacht, Germany}
}

\date{}

\maketitle

\begin{abstract}
Cardiac growth and remodeling (G\&R) patterns change ventricular size, shape, and function both globally and locally. Biomechanical, neurohormonal, and genetic stimuli drive these patterns through changes in myocyte dimension and fibrosis. We propose a novel microstructure-mo\-ti\-vat\-ed model that predicts organ-scale G\&R in the heart based on the homo\-genized con\-strained mixture theory. Previous models, based on the ki\-ne\-ma\-tic growth theory, reproduced con\-se\-quenc\-es of G\&R in bulk myocardial tissue by prescribing the direction and extent of growth but neglected underlying cellular mechanisms. In our model, the direction and extent of G\&R emerge naturally from intra- and extracellular turnover processes in myocardial tissue constituents and their preferred homeostatic stretch state. We additionally propose a method to obtain a mechanobiologically equilibrated reference configuration. We test our model on an idealized 3D left ventricular geometry and demonstrate that our model aims to maintain tensional homeostasis in hypertension conditions. In a stability map, we identify regions of stable and unstable G\&R from an identical parameter set with varying systolic pressures and growth factors. Furthermore, we show the extent of G\&R reversal after returning the  systolic pressure to baseline following stage 1 and 2 hypertension. A realistic model of organ-scale cardiac G\&R has the potential to identify patients at risk of heart failure, enable personalized cardiac therapies, and facilitate the optimal design of medical devices.
    
    \keywords{Cardiac growth and remodeling \and Homogenized constrained mixture model
    \and Computational modeling \and Mechanobiology \and Hypertension}
\end{abstract}

\section{Introduction}
\label{intro}

Cardiac growth and remodeling (G\&R) occur in various situations throughout a human's life. Developmental growth occurs between
birth and adulthood, and ensures the adaption of the cardiac function to the changing needs. Changing
needs also occur in adulthood, for example, due to pregnancy or exercise resulting in physiological
growth of the myocardium. Disease-induced stimuli, however, often result in pathologic, maladaptive
G\&R that develop towards heart failure. Common diseases stimulating pathologic G\&R are
myocardial infarction, aortic stenosis, hypertension, or valvular regurgitation \citep{Cohn2000a}.
This paper will focus on pathological G\&R in the adult heart.

\subsection{Phenotypes of cardiac G\&R in heart failure}

Two measures are commonly used to characterize different cardiac G\&R patterns: The mass of the left
ventricle (LV) and the ratio of LV wall thickness to the diastolic diameter, denoted as relative
wall thickness. With these two measures, cardiac G\&R can be classified into two distinct patterns:

If both the relative wall thickness and LV mass is increased, the pattern is classified
as \emph{concentric hypertrophy}. This pattern is typically observed in patients with
increased resistance to ejection, e.\,g. aortic stenosis or hypertension. On the cellular
level, contractile-protein units (sarcomeres) in cardiomyocytes are added in parallel so that the length to width
ratio of individual cardiomyocytes decreases \citep{Kehat2010a}.

An increase in LV mass with constant relative wall thickness is classified
as \emph{eccentric hypertrophy}. This pattern is typically a result of valvular insufficiencies \citep{Linzbach1960a}. 
Cardiomyocytes increase their length to width ratio by serial addition of
sarcomeres \citep{Kehat2010a}.

Ventricular and cellular G\&R in the heart also involves changes in the collagen network that surround
each myocyte \citep{Kehat2010a}. Increased collagen fraction in the myocardium, i.\,e. myocardial fibrosis,
results in a higher stiffness and causes poor diastolic filling characteristics \citep{Spinale2007a},
increasing the risk of heart failure \citep{Brower2006a}. Additionally, fibrosis impairs contractility and electrophysiology,
causing arrhythmias, local microfibrillations, and inefficient contraction \citep{Schirone2017a}. Increased
collagen deposition is mainly observed in pressure over\-load\-ed cases. In contrast, a loss of collagen
fibrils is observed in cases with volume overload \citep{Spinale2007a}, supporting the dilatory
process of the ventricle \citep{Kehat2010a}.

\subsection{Fundamentals of G\&R}
\label{sec:into.fundamentals}
We will only briefly discuss the mechanobiological processes behind G\&R of soft
tissue. The interested reader is referred to the review of \cite{Cyron2017a} and
references therein.

\subsubsection{Homeostasis}

It is generally accepted that there is a preferred mechanical state within living tissue that mechanobiological activity
seeks to retain \citep{Eichinger2021b}. This state is often referred to as homeostatic state.

In vitro studies with collagen-gels seeded with fibroblasts support this hypothesis. The
initially stress-free gels develop internal stresses that reach a plateau after a few hours. A
subsequent stretching or relaxing of the gels, i.e., a perturbation from the apparently homeostatic
state, immediately changes the stress level followed by a slow, exponential return back toward
the homeostatic plateau \citep{Eichinger2021a,Cyron2017a,Brown1998a, Ezra2010a}.

Cardiomyocytes are known to react to increased loading conditions by synthesizing new contractile proteins
and assembling new sarcomeres \citep{Cohn2000a} to adapt their size.
\cite{Yang2016a} have shown in \emph{in vitro} experiments that new sarcomeres were assembled within the cell.
Although in vitro studies of homeostasis of cardiomyocytes are rare, it is plausible that
also cardiomyocytes have some preferred mechanical state that they try to maintain. Given their high volume
fraction and their importance during systole, \cite{Grossman1975a} support
this with their broadly accepted wall stress hypothesis that states that myocardial wall thickness increases to return
systolic stress to normal.

\subsubsection{Turnover}

Turnover is a key underlying process by which G\&R occurs in living soft tissue. It is the continuous
deposition and active degradation of tissue constituents \citep{humphrey2002a}.
In the physiological case, it maintains tissue integrity and prevents mechanical fatigue. The amount
of deposited (secreted) mass is balanced with the actively degraded mass so that overall tissue
mass and structure do not change (tissue maintenance). In case of injury or disease, there can be
imbalance resulting in effective increase or decrease of tissue mass to return constituent stresses
back to homeostasis.

Turnover rates of cardiomyocytes are low: Fewer than $50\,\%$ of cardiomyocytes are replaced during
lifetime in a healthy heart \citep{Bergmann2009a}. Less than $1\,\%$ of cardiomyocytes in a healthy adult
heart are replaced per year \citep{Bergmann2015a}. However, to maintain cardiac function, there exists an orchestrated
intracellular process that synthesizes, assembles and degrades proteins of the sarcomeres inside cardiomyocytes
\citep{Willis2009a}. Turnover in the context of cardiomyocytes can therefore be seen as an
intracellular process \citep{humphrey2002a}.

The collagen network that surrounds cardiomyocytes also turnover. Around $0.6\,\%$ of the collagen is synthesized per day in the healthy state \citep{Weber1989a} by fibroblasts \citep{Humphrey2002b}. Their half-life time is estimated to be $80$ to $120\,\si{days}$ \citep{Weber1989a}. In case of injury, a six- to eight-fold increase in collagen synthesis is reported \citep{Weber1989a}.

\subsection{Mathematical models of cardiac G\&R}
\label{sec:existingmathematicalmodels}

Many different mathematical models have been proposed so far in the field of cardiac
G\&R. We only discuss a few of them. The interested reader is instead referred to the review
papers of \cite{Lee2016b}, \cite{Aboelkassem2019a}, \cite{Yoshida2020a}, \cite{Niestrawska2020a},
and \cite{Sharifi2021a}.

An important group of models originates from the kinematic growth theory introduced by \cite{Rodriguez1994a}.
They propose a multiplicative split of the deformation gradient into an inelastic, growth-related part
and an elastic part. \cite{Kroon2009a} used this model to simulate inhomogeneous 3D
isotropic volumetric growth of a truncated ellipsoid. Since then, the model has been modified and extended several times.
For example, \cite{Goektepe2010a} simulated concentric and eccentric growth, \cite{Lee2015a} extended the model to capture also
reversal of growth. This model has also been used to reconstruct growth features from in vivo MRI \citep{Fan2021a}
and has been coupled with 0D models of stimuli from hormonal signals \citep{Estrada2020a}.

As \cite{humphrey2002a} pointed out, kinematic growth theory mainly models certain consequences of growth.
It captures the outcome of G\&R only phenomenologically \citep{Niestrawska2020a} instead of modeling the
fundamental ongoing processes in living tissue. For example, such models limit
growth by artificially introducing a maximal pathological myocyte inelastic stretch or limiting the
number of growth steps. In a predictive model of cardiac G\&R, however, mechanobiological stability \citep{Cyron2014a,Cyron2014b} of the G\&R process must be a result of the
geometry, tissue properties, and severity of the pathological event, rather than of a priori assumptions. As \cite{Yoshida2020a}
point out, simple kinematic growth models also often fail to predict reverse growth that is expected after the removal of the
pathological load (e.\,g., repair of an aortic stenosis).

A fundamentally different approach has been proposed by \cite{humphrey2002a}. They model soft
tissue as a mixture of different constituents. Mass increments of constituents are deposited
at every point in time into the mixture and existing mass is degraded over time -- as observed
in the mechanobiology of tissue turnover. Hence, G\&R is a consequence of modeling fundamental
processes observed in living tissue. The model, so far, has mostly been applied to vascular G\&R. Due to the computational costs
of the model evaluation and the complexity of the model, 3D models of G\&R are rare. \cite{cyron2016a} proposed a temporal homogenization
of all mass depositions for each constituent resulting in a drastic reduction of the computational costs.
The so-called homogenized constrained mixture model has been applied to 3D vascular G\&R on a thick-walled cylinder
\citep{braeu2016a} and patient-specific geometries \citep{Mousavi2019a}. To the best of our knowledge, constrained mixture models
have not been applied to cardiac G\&R yet.

\subsection{Outline}

In this paper, we propose a homogenized constrained mixture model for cardiac G\&R that
bases on the principal mech\-anobiological processes described above. We demonstrate the capabilities
of the model on a truncated ellipsoid model of an idealized LV to capture
mechanobiological stability and reversal of cardiac G\&R. Finally, we discuss the results and outline possible further steps for improvement.

\section{Mathematical Modeling}
\label{sec:modeling}

Let $\mathcal{B}_0$ be the reference configuration of a body at time\footnote{We use $s$ for the G\&R time to emphasize the difference to the timescale of a heartbeat usually denoted as $t$.} $s_0=0$ and $\mathcal{B}_s$ the
configuration at time $s$. A material point $\vct{X} \in \mathcal{B}_0$ is mapped to its current
position $\vct{x} \in \mathcal{B}_s$ via
\begin{align*}
    \vct{x}: \mathcal{B}_0 \times [0, \infty) \rightarrow \mathcal{B}_s, \quad (\vct{X}, s) \mapsto \vct{x}(\vct{X}, s).
\end{align*}
The displacement field is $\vct{u}=\vct{x}-\vct{X}$ and the deformation
gradient is $\mat{F} = \frac{\pd \vct{x}}{\pd \vct{X}}$ with the Jacobian
determinant $J = \det \mat{F}$.

G\&R occurs on very long time scales so that inertial effects can be neglected. Using the principle of virtual work \citep{holzapfel2000a}, the static mechanical equilibrium can be written as
\begin{align}
    \delta W = \int_{\mathcal{B}_0} \mat{P} : \delta \mat{F} \, \dd V = 0,
    \label{eqn:principle_of_virtual_work}
\end{align}
where $\mat{P}$ is the first Piola-Kirchhoff stress tensor.

\subsection{Microstructural modeling of the myocardium}
\label{sec:math:microstruc}

We model the myocardium as a constrained mixture that consists of multiple structurally relevant
constituents. All constituents $i$ occupy all points $\vct{x} \in \mathcal{B}_s$ and 
deform together with no relative motion between them ($\mat{F} = \mat{F}^i$). Quantities
related to a specific constituent are denoted by superscript $i$.

G\&R is modeled by a homogenized constrained mixture model as proposed by \cite{cyron2016a} and \cite{braeu2016a,braeu2019a}. In contrast to
classical constrained mixture models \citep{humphrey2002a}, the deposition and degradation of mass
increments is captured only in a temporally homogenized sense. Instead of capturing the natural
configuration of every mass increment, only an averaged inelastic deformation of each constituent
$\mat{F}_\text{r}^i$ is stored. The total deformation can be written as
\begin{align}
    \mat{F} = \mat{F}_\text{e}^i \mat{F}_\text{r}^i,
    \label{eqn:def_grd_split}
\end{align}
where $\mat{F}_\text{e}^i$ is the elastic deformation of constituent $i$ that ensures that (\ref{eqn:principle_of_virtual_work})
is satisfied and all constituents deform together.

The first Piola-Kirchhoff stress tensor $\mat{P}$ in (\ref{eqn:principle_of_virtual_work}) is
\begin{align}
    \mat{P} = \frac{\pd \Psi}{\pd \mat{F}},
    \label{eqn:stress_response}
\end{align}
with the strain energy (per unit volume)
\begin{align}
    \Psi = \sum_{i=0}^n \rho_0 ^i W^i \left(\mat{C}_\text{e}^i\right) + \Psi^\#.
    \label{eqn:total_strain_energy}
\end{align}
$W^i$ is the strain energy per unit mass, $\rho_0^i$ is the reference mass density of constituent $i$, and
$\mat{C}_\text{e}^i = \mat{F}_\text{e}^{i\T} \mat{F}_\text{e}^i$ is the Cauchy-Green
deformation tensor of the elastic deformation of constituent $i$.
This homogenization of the stress response across all constituents is denoted as
\emph{rule-of-mixtures} \citep{humphrey2002a}. Only the equilibrium for the whole mixture has to be solved and not for each
constituent individually. $\Psi^\#$ is a penalty-type strain energy function per unit volume to ensure
a (nearly) constant spatial density of the whole tissue at any time (incompressibility).

The strain energy functions of the constituents are only functions of the constituent's elastic deformation.
Deformation that stems from the inelastic deformation (i.e., G\&R) does not store elastic energy and
therefore does not directly create stresses.

We assume that the primarily structurally relevant constituents of the myocardium are collagen fibers ($i=\text{c}$)
and cardiomyocytes ($i=\text{m}$). Both constituents are modeled to be quasi-1D constituents that only bear
stresses in their preferred direction. The strain energy per unit mass of the collagen fibers is an exponential of the form
\begin{align}
    W^\text{c} = \frac{a^\text{c}}{2b^\text{c}} \left\{ \exp \left[b^\text{c} \left(I_{4}^\text{c}-1\right)^2\right] -1 \right\},
    \label{eqn:strain_energy_collagen_fibers}
\end{align}
where $I_{4}^\text{c} = \vct{f}_r^\text{c} \cdot \mat{C}_\text{e}^\text{c} \vct{f}_r^\text{c} $ is the fourth pseudo-invariant of 
$\mat{C}_\text{e}^\text{c}$ \citep{holzapfel2000a} and $\vct{f}_r^\text{c}$ is the preferred direction of the collagen fiber family
in the intermediate configuration (related to the reference configuration via $\mat{F}_\text{r}^\text{c}$).

For cardiomyocytes, we use an additively decomposed strain energy function 
\begin{align}
    W^\text{m} = W_\text{pas}^\text{m} + W_\text{act}^\text{m},
    \label{eqn:strain_energy_myocyte_fibers}
\end{align}
to incorporate passive and active contributions, respectively. The passive contribution is modeled with an exponential of the form
\begin{align}
    W_\text{pas}^\text{m} = \frac{a^\text{m}}{2b^\text{m}} \left\{ \exp \left[b^\text{m} \left(I_{4}^\text{m}-1\right)^2\right] -1 \right\},
    \label{eqn:strain_energy_fibers}
\end{align}
analogously to collagen fibers.

Active muscle tone has not been developed yet for cardiomyocytes in the G\&R timescale. We,
therefore, adopt a simple model from \cite{Wilson2012a} for vascular smooth muscle tone via
an additional term in the strain energy function:
\begin{align}
    W_\text{act}^\text{m} = \frac{\sigma_{\text{act}}^\text{m}}{\rho_0(s=0)} \left( \lambda_\text{act}^\text{m} + \frac{1}{3} \frac{(\lambda_\text{max}^\text{m}-\lambda_\text{act}^\text{m})^3}{(\lambda_\text{max}^\text{m}-\lambda_0^\text{m})^2} \right).
    \label{eqn:strain_energy_myocyte_active}
\end{align}
$\sigma_{\text{act}}^\text{m}$ is the maximal active Cauchy stress, $\lambda_\text{act}^\text{m}$ is the active stretch
in fiber direction, and $\lambda_0^\text{m}$ and $\lambda_\text{max}^\text{m}$ are the active stretch at zero and
maximum active stress. In accordance to \cite{Wilson2012a}, we assume that the active muscle tone evolves with the
G\&R of cardiomyocytes, such that $\frac{\partial \lambda_{\text{act}}^\text{m}}{\partial \lambda^{\text{m}}} = \frac{1}{\lambda^\text{m}}$,
where $\lambda^\text{m}$ is the current total stretch of the muscle fiber compared to the reference configuration \citep{braeu2019a}.

\reva{
    Collagen fibers and cardiomyocytes are excluded during compression\footnote{\reva{Note that no fibers are under compression in the examples within the paper.}} ($I_4^i<1$) since their contribution to compressive stresses is usually assumed to be negligible \citep{Holzapfel2009a}.
}

The remaining structural constituents (mainly elastin; $i=\text{3D}$) are modeled with a decoupled isotropic neo-Hookean
strain energy contribution:
\begin{align}
    W^\text{3D} = c_1 \left(\bar{I}_1-3\right).
    \label{eqn:strain_energy_elastin}
\end{align}
$\bar{I}_1 = J^{-\nicefrac{2}{3}} \tr{\left(\mat{G}^{\text{3D}} \mat{F}^\T \mat{F} \mat{G}^{\text{3D}}\right)}$ is the first modified invariant \citep{holzapfel2000a} of the total elastic
Cauchy-Green deformation tensor of elastin including the isochoric and rotation-free prestretch tensor of elastin $\mat{G}^\text{3D}$. The prestretch tensor maps the stress-free configuration of elastin into the reference configuration.

\reva{In the following, we will introduce the needed evolution equations for G\&R. Our model currently
does not consider direct interactions between constituents, hence, we can formulate these evolution equations for each constituent
separately.}

\subsection{Mass production and removal}

As described earlier, the continuous deposition and degradation of mass is called tissue turnover.
Turnover is modeled as a superposition of deposition and degradation of mass. We therefore can
split the net mass production rate $\dot{\rho}_0^i$ into a true mass production rate $\dot{\rho}_{0+}^i$
and a true mass degradation rate $\dot{\rho}_{0-}^i$:
\begin{align*}
    \dot{\rho}_0^i = \dot{\rho}_{0+}^i + \dot{\rho}_{0-}^i.
\end{align*}
In homeostasis, the deposition exactly compensates for the degradation of mass. We track the deposition and degradation
of mass by an evolution of the mass density in the \emph{reference} configuration. Hence, the amount of mass in a
reference volume element changes over time, and mass is conserved between the evolved reference configuration and the
current configuration.

We assume the following constitutive relation for the net mass production rate:
\begin{align}
    \dot{\rho}_{0}^i = \rho_{0}^i k_\sigma^i \frac{\sigma^i-\sigma_\text{h}^i}{\sigma_\text{h}^i},
    \label{eqn:net_mass_production_rate}
\end{align}
where $\sigma^i$ is the current Cauchy stress of constituent $i$ in their preferred direction, $\sigma_\text{h}^i$ is the
preferred homeostatic stress state and $k_\sigma^i$ is a growth gain factor.

The degradation of mass is assumed to follow a simple Poisson process
\begin{align}
    \dot{\rho}_{0-}^i (s) = -\frac{\rho_0^i(s)}{T^i},
    \label{eqn:mass_degradation}
\end{align}
where $T^i$ is the mean lifetime of a deposited mass increment.

\subsection{Remodeling}
\label{sec:modeling:remodeling}

\begin{figure*}
    \begin{center}
        \import{figures/}{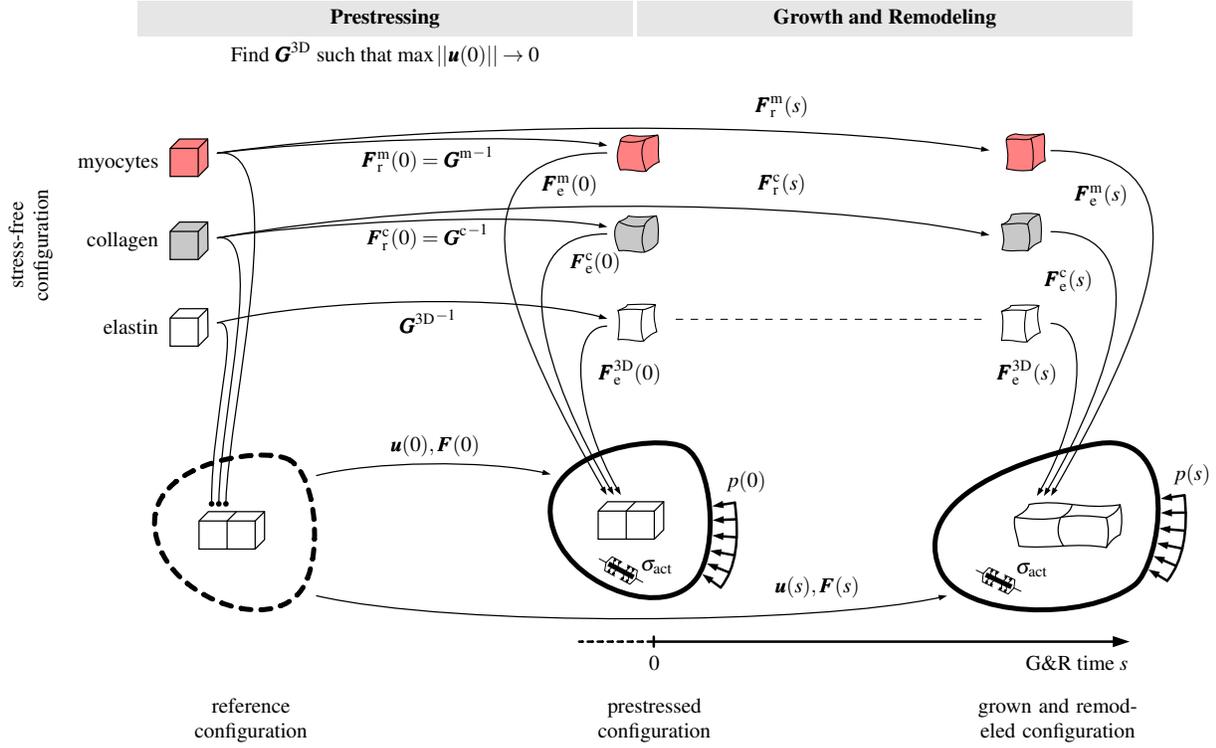}
    \end{center}
    \caption{The different configurations during prestressing, and growth and remodeling.
    Each configuration consists of multiple constituents, namely cardiomyocytes, collagen fibers
    and elastin, which occupy every point in the domain at once. The reference configuration is
    obtained from imaging data of a heart in systole. The initial condition for the remodeling
    deformation gradients of cardiomyocytes and collagen fibers are chosen such that these constituents
    are in their preferred mechanical environment in the reference ($\widehat{=}$ imaged) configuration.
    Prestress, active stress and external loading cause the mixture to elastically deform.
    The prestress algorithm aims at an elastin prestretch tensor $\mat{G}^\text{3D}$ such that the
    prestressed and the reference configurations are identical ($\vct{u}(0)=\vct{0}$), or at least similar.
    The prestressed configuration, therefore, is in (or close to) mechanobiological equilibrium. A
    subsequent pathological event (e.g. an increase of the pressure $p$) triggers growth and remodeling of the tissue. The stress-free
    configurations of cardiomyocytes and collagen fibers evolve through the continuous
    deposition and degradation of mass increments that are captured in a homogenized sense in $\mat{F}_\text{r}^{i}(s)$.}
    \label{fig:configurations}
\end{figure*}

Remodeling of the tissue is a direct consequence of continual turnover. 
At every point in time, new mass increments of cardiomyocytes and collagen fibers are deposited and
extant constituents continuously degrade. New mass increments of constituent $i$ deposit at rate $\dot{\rho}_{0+}^i$
and are assembled in general in a different stress state than the extant material of the constituent. Extant
mass is continuously degraded at a rate of $\dot{\rho}_{0-}^i$. This deposition and degradation of mass
locally change the stress state of the constituent, resulting in a rearrangement of the constituent
which is remodeling of the tissue.

Homogenized constrained mixture models capture the mass increments and their stress-free configurations
in a temporally homogenized sense by an adaption of the stress-free configuration of the constituent (see Figure~\ref{fig:configurations}).
We assume herein that new mass increments are deposited and assembled in their preferred mechanical state
$\mat{\sigma}_{\text{h}}^i$ and extant mass of the constituent is continuously degraded with the current Cauchy stress $\mat{\sigma}^i$.
\cite{cyron2016a} and \cite{braeu2016a} have shown that this change in the stress-free configuration can
be described by an evolution of an inelastic part of the deformation gradient with the equation
\begin{align}
    \left[
        \frac{\pd \mat{\sigma}^i}{\pd \mat{F}_\text{e}^i} : \left(\mat{F}_\text{e}^i \mat{L}_\text{r}^i\right)
        \right]_{\mat{F}=\text{const.}} = \left(
    \frac{\dot{\rho}_0^i (s)}{\rho_0^i} + \frac{1}{T^i}
    \right) \left(\mat{\sigma}^i-\mat{\sigma}_{\text{h}}^i\right),
    \label{eqn:remodeling_main_equation}
\end{align}
where $\mat{L}_\text{r}^i = \dot{\mat{F}}_\text{r}^i {(\mat{F}_\text{r}^i)}^{-1}$ is the remodeling velocity gradient.
The stiffness $\frac{\pd \mat{\sigma}^i}{\pd \mat{F}_\text{e}^i}$ for the quasi one-dimensional
fiber families as defined in Equation~(\ref{eqn:strain_energy_fibers}) is rank deficient. Hence,
the remodeling deformation modes that do not contribute to strain energy can be chosen arbitrarily. If we
assume incompressible remodeling and that newly deposited fibers are always aligned in the same direction in the reference
configuration, the rotational part of $\mat{F}_\text{r}^i$ is the identity matrix. We can then
write \citep{cyron2016a}
\begin{align}
    \mat{F}_\text{r}^i = \lambda_\text{r}^i \vct{f}_0^i \otimes \vct{f}_0^i + \frac{1}{\sqrt{\lambda_\text{r}^i}}
    (\mat{I}-\vct{f}_0^i \otimes \vct{f}_0^i),
    \label{eqn:remodeling_deformation_gradient}
\end{align}
where the inelastic remodeling stretch in fiber direction $\lambda_\text{r}^i$ is the only unknown. A
consequence of the rotation-free $\mat{F}_\text{r}^i$ is that we can use $\vct{f}_0^i$ which is the unit
vector field in the \emph{reference} configuration of the preferred direction of the fiber constituent.
Equation~(\ref{eqn:remodeling_main_equation}) can then be simplified so that the only unknown
$\lambda_\text{r}^i$ can be computed with (see Appendix 1 in \cite{cyron2016a} for details of the
derivation)
\begin{align}
    \label{eqn:remodeling_evolution_equation_oned}
    \dot{\lambda}_\text{r}^i = \left[\frac{\dot{\rho}_0^i}{\rho_0^i} + \frac{1}{T^i}\right]
    \frac{\lambda_r^i}{2\ I_4^i} \left[\frac{\pd \sigma^i}{\pd I_4^i}\right]^{-1}\left(\sigma^i - \sigma_\text{h}^i\right).
\end{align}
Equation~(\ref{eqn:remodeling_evolution_equation_oned}) can be integrated in time at every
Gauss point using a standard time integration scheme starting from an initial condition. As simulations are often started from a homeostatic state, the initial remodeling stretch is chosen such that the elastic stretch of the constituent initially equals the homeostatic stretch. Using equation (\ref{eqn:def_grd_split}) and $\mat{F}(s=0)=\mat{I}$ (identity matrix), we get
\begin{align*}
    \lambda_\text{r}(s=0) = \frac{1}{\lambda_\text{h}^i}.
\end{align*}

\begin{figure}
    \begin{center}
        \begin{tikzpicture}

    \draw (0,0) coordinate(bottom) -- ++(0,0.5) -- ++(0.5,0,0) --
    ++(0,0.25) coordinate (spring_bottom) -- ++(0.25,0.1) -- ++(-0.5,0.2) -- ++(0.5,0.2) -- ++(-0.5,0.2) ++(0.5,0) node[right]{$C^i$} ++(-0.5,0) -- ++(0.5,0.2) -- ++(-0.5,0.2) -- ++(0.5,0.2) -- ++(-0.25,0.1) --
    ++(0,0.25) coordinate (spring_top) coordinate (dashpot_bottom) --
    ++(0.25,0) -- ++(0,0.25) node[right] {$\frac{\dot{\rho}_0^i}{\rho_0^i}+\frac{1}{T^i}$} -- ++(0,0.25) (dashpot_bottom) -- ++(-0.25,0) -- ++(0,0.5) (dashpot_bottom) ++(0,0.25) coordinate(dashpot_top) -- ++(0.2,0) (dashpot_top) -- ++(-0.2,0) (dashpot_top) -- ++(0,0.5) --
    ++(-0.5,0) coordinate(crossing_top) -- ++(0,0.5) coordinate(top) (crossing_top) -- ++(-0.5,0) --
    ++(0,-2.65/2+0.1) ++(0,-0.1) circle (0.1) ++(-0.1,0) node[left] {$\sigma_\text{h}^i$} ++(0.1,0) ++(0,-0.1) -- ++(0,-2.65/2+0.1) -- ++(0.5,0);

    \fill (bottom) circle(0.1) ++(0,-0.1) node[below] {$\sigma^i$};
    \fill (top) circle(0.1) ++(0,0.1) node[above] {$\sigma^i$};
\end{tikzpicture}
    \end{center}
    \caption{
        The physical analog model of a homogenized constrained mixture model subjected to turnover
        as shown by \cite{cyron2016a}. A Maxwell element (right) with a spring with stiffness $C^i$ in
        and dashpot with time constant $\frac{\dot{\rho}_0^i}{\rho_0^i}+\frac{1}{T^i}$ parallel to a
        motor unit (left) exerting the homeostatic (pre-)stress $\sigma_\text{h}^i$.
    }
    \label{fig:maxwell_motor_element}
\end{figure}
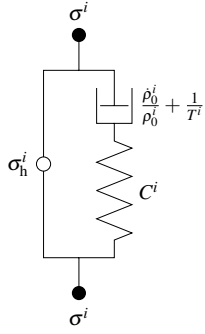

\cite{cyron2016a} derived a physical interpretation of turnover in homogenized constrained mixture
models. They showed that a Maxwell element with a parallel motor unit as shown in
Figure~\ref{fig:maxwell_motor_element} is a mechanical analog model. The Maxwell element is a
spring with some nonlinear stiffness and a dashpot with time constant $\frac{\dot{\rho}_0^i}{\rho_0^i}+\frac{1}{T^i}$
in series. The motor element exerts the homeostatic stress $\sigma_\text{h}^i$. When the current
stress state deviates from the homeostatic stress state of the constituent, the deviation has to be
borne by the Maxwell element. Maxwell elements are inherently unable to support static loads resulting
in an isochoric shift of the stress-free length of the constituent.

In the kinematic growth theory, G\&R is captured by defining an inelastic G\&R tensor phenomenologically. Here,
the remodeling deformation gradient is derived from the ideas of the full constrained mixture by temporal
homogenization of mass turnover. This model is mechanobiologically motivated with computational costs
comparable to the kinematic growth theory.

\subsection{Growth}

By now, we have modeled continual turnover of the constituents and how it causes an inelastic (isochoric) remodeling. Still missing is the modeling of the volume change due to mass increase or decrease, that is, growth.

There are different approaches that describe the volume change due to mass growth. An often chosen approach is to
define an additional inelastic deformation $\mat{F}_\text{g}^i$ for each constituent that incorporates the volume change of the tissue. The definition of such an inelastic deformation
typically requires the definition of a structural tensor that defines the anisotropy of growth. As obtaining this information from imaging data is difficult, anisotropy of growth is often chosen such that the obtained results are reasonable. This hinders the approach from being truly predictive.

\cite{braeu2019a} proposed a way that bypasses the controversial definition of the
growth tensor. They assumed that the deposition of new mass results in an elastic distention
of whole differential volume elements also affecting other constituents. Due to the high stiffness of collagen fibers and cardiomyocytes, this elastic distention mainly occurs perpendicular to the existing fibers. Remodeling according to (\ref{eqn:remodeling_main_equation}) transforms this (in general) anisotropic distension over time into an inelastic deformation.

We are following this approach and modify the penalty type strain energy term $\Psi^\#$
that approximates incompressibility so that it nearly ensures a constant spatial density:

\begin{align}
    \Psi^\# = \frac{\kappa}{2} \left(|\mat{F}|-\frac{\rho_0(s)}{\rho_{0}(s=0)}\right)^2.
    \label{eqn:const_spatial_mass_density_penalty}
\end{align}
The current reference mass density $\rho_0$ of the whole tissue is
\begin{align*}
    \rho_0(s) = \sum_{i=0}^n \rho_0^i(s).
\end{align*}

When choosing the penalty parameter $\kappa$ sufficiently high, a nearly constant
spatial density is ensured. The minimization of the total strain energy will then
result in a dilation mainly in the direction with the smallest stiffness. The interested reader is referred to \cite{braeu2019a}, who
theoretically analyzed the natural anisotropic growth stemming from anisotropic
stiffness of the tissue.

\subsection{Prestress}
\label{sec:modeling:prestress}

The geometry for a G\&R simulation is usually a configuration obtained from
Magnetic Resonance Imaging (MRI) through segmentation. External loads present
during imaging, e.\,g, from the blood pressure $p$ and the surrounding tissue,
result in an in vivo reference configuration that is not stress-free. Since we model the
tissue as a constrained mixture, each constituent can have its own (incompatible)
stress-free configuration. Note that it is necessary that the in vivo configuration is not stress-free such that all constituents
can be in their homeostatic state for a given left ventricular pressure $p$.

In our examples, we assume that the imaged configuration represents a healthy state before some physiological event occurs
that stimulates G\&R. The configuration must therefore be in mechanobiological equilibrium, which requires
mechanical equilibrium and additionally that all constituents are in their preferred homeostatic mechanical state \citep{Cyron2014a}.
As cardiomyocytes occupy the largest volume fraction in cardiac tissue, we motivate the selection
of the reference configuration based on them. We heuristically assume that cardiomyocytes attain their homeostatic state during systole, where their generated
active force is at maximum. Note that tensional homeostasis of tissue under pronounced cyclical loading remains poorly understood to date
and the definition of such a configuration is part of ongoing research.

The constituents subject to G\&R are the cardiomyocytes and the collagen fibers. Both are
quasi-1D fiber constituents. They are assumed to be in their homeostatic state if their current stress state equals their
preferred stress state ($\sigma^i = \sigma_\text{h}^i$) that is known and a-priori defined. We will
define this preferred state using the homeostatic stretch $\lambda_\text{h}^i$ obtaining
the homeostatic stress from the constitutive relation.

\cite{Cocciolone2018a} found that synthesis of functional elastin in vessels mainly happens during development
and maturation. Lacking of specific experiments, we assume similar behavior in the myocardium.
Our assumption is that elastin does not have a mechanism to maintain
its preferred stress state in maturity.  Therefore, even in mechanobiological equilibrium, the stress response of elastin is non-ho\-mo\-geneous in contrast to the stress responses of cardiomyocytes and collagen fibers. Figure~\ref{fig:configurations} depicts the different configurations during prestressing. The prestress algorithm aims for the elastin prestretch tensor $\mat{G}^\text{3D}$ such that the prestressed configuration is identical (or at least similar) to the reference ($\widehat{=}$ imaged) configuration. The prestretch tensor of collagen fibers and cardiomyocytes are given by
\begin{align*}
    \mat{G}^i = \lambda_\text{h}^i \vct{f}_0^i \otimes \vct{f}_0^i + \frac{1}{\sqrt{\lambda_\text{h}^i}} \left( \mat{I} - \vct{f}_0^i \otimes \vct{f}_0^i \right),
\end{align*}
such that these constituents are in their preferred mechanical state in the reference configuration

\cite{mousavi2017a} and \cite{weisbecker2014a} introduced an iterative algorithm that we will adopt
with a modification: Since only the deviatoric stretches contribute to the internal energy of elastin,
we choose $\mat{G}^\text{3D}$ to be isochoric and rotation-free (symmetric). 

To ease the solution of the nonlinear equilibrium equations, the external forces and the a-priori defined prestretch of the fibrillar constituents are applied gradually in the first couple of timesteps. Our initial guess is that elastin is stress-free in the reference configuration, hence $\mat{G}_{0}^\text{3D} = \mat{I}$.
After every prestress iteration $k$, we compute the new isochoric prestretch deformation gradient
\begin{align*}
    \bar{\mat{F}}_{k+1,\text{pre}} = \bar{\mat{F}} \mat{G}_k^\text{3D},
\end{align*}
with $\bar{\mat{F}} = \left(J\right)^{-\frac{1}{3}} \mat{F}$ being the isochoric part
of the deformation gradient. To ensure that the prestretch tensor is rotation-free, a polar
decomposition is computed yielding the new prestretch tensor $\mat{G}_{k+1}^\text{3D}$:
\begin{align*}
    \mat{G}_{k+1}^\text{3D} = \mat{R}_{k+1,\text{pre}}^\T \bar{\mat{F}}_{k+1,\text{pre}},
\end{align*}
where $\mat{R}_{k+1,\text{pre}}$ is the rotational part of $\bar{\mat{F}}_{k+1,\text{pre}}$.

This iterative algorithm is repeated until the maximum Euclidean norm of the nodal displacements between the reference and the prestressed configuration falls below a tolerance $\varepsilon_\text{pre}$. This tolerance ensures that the maximum deviation of the prestressed configuration from the reference ($\widehat{=}$ imaged) configuration is bounded. It can be chosen depending on the resolution of the imaging data and the acceptable difference between these configurations. It is important to note that the displacements after prestressing also influence the mechanobiological equilibrium of the 1D constiuents. A subsequent phase of G\&R can achieve mechanobiological equilibrium at the cost of additional displacements and, therefore, a further deviation from the reference configuration.

\subsection{Solving the G\&R problem}
\label{sec:modeling:implementation}

The resulting model of G\&R is a quasi-static problem given by equation~(\ref{eqn:principle_of_virtual_work}) with local
evolution equations given by equations~(\ref{eqn:net_mass_production_rate}) and (\ref{eqn:remodeling_evolution_equation_oned})
on each integration point. The proposed model can be easily implemented in many existing computational
frameworks of classical solid mechanics as a constitutive model. To compute the stress given
the deformation gradient of the mixture at each integration point, the following steps have to be done:
\begin{enumerate}
    \item Integrate the local evolution equations (\ref{eqn:net_mass_production_rate}) and (\ref{eqn:remodeling_evolution_equation_oned}) using a standard time integration scheme.
    \item Compute the remodeling deformation gradient of each constituent using equation~(\ref{eqn:remodeling_deformation_gradient}).
    \item Compute the elastic part of the deformation of each constituent by equation~(\ref{eqn:def_grd_split}).
    \item Compute the stress response of the mixture with equations~(\ref{eqn:stress_response}) - (\ref{eqn:strain_energy_elastin}) and (\ref{eqn:const_spatial_mass_density_penalty}).
\end{enumerate}
The nonlinear equilibrium equation~(\ref{eqn:principle_of_virtual_work}) is solved with a
New\-ton-Raphson type algorithm. Within that, the steps above have to be done in every iteration. If an
explicit time integration is used in step 1, steps 1 and 2 have to be done only once per timestep.

\section{Numerical examples}
\label{sec:results}

\begin{figure}
    \begin{center}
        \begin{tikzpicture}
    \pgfmathsetmacro{\aendo }{2}
    \pgfmathsetmacro{\bendo }{1}
    \pgfmathsetmacro{\aepi }{2.5}
    \pgfmathsetmacro{\bepi }{1.5}
    \pgfmathsetmacro{\c }{1}
    \pgfmathsetmacro{\eps }{0.1}
    \pgfmathsetmacro{\measx }{0.7}

    \newcommand{\spring}[2][0]
    {
        \pgfmathsetmacro{\springwidth}{0.5}
        \pgfmathsetmacro{\springheight}{0.2}
        \pgfmathsetmacro{\springnum}{5}
        \pgfmathsetmacro{\springeps}{0.05}

        \pgfmathsetmacro{\springdx}{(\springwidth-2*\springeps)/(2*\springnum+1)}

        \begin{scope}[shift={(#2)}]
            \begin{scope}[rotate=#1]
                \draw (0,0) -- ++(\springeps,0) -- ++(\springdx/2,-\springheight/2)
                    % foreach
                    \foreach \tmp in {1,...,\springnum}
                    {-- ++(\springdx,\springheight) -- ++(\springdx,-\springheight)}
                    -- ++(\springdx/2,\springheight/2) -- ++(\springeps,0);
            \end{scope}
        \end{scope}
    }
    \newcommand{\omnispring}[2][0]
    {
        \begin{scope}[rotate around={#1:(#2)}]
            \spring[0]{#2}
            \begin{scope}[shift={(0.5,0)}]
                \wall[90]{#2}
            \end{scope}
        \end{scope}
    }
    \newcommand{\normalspring}[2][0]
    {
        \begin{scope}[rotate around={#1:(#2)}]
            \fill[white] (#2)+(0,-0.2) rectangle ++(0.65,0.2);
            \spring[0]{#2}
            \begin{scope}[shift={(0.5,0)}]
                \freewall[90]{#2}
            \end{scope}
        \end{scope}
    }
    
    \pgfmathsetmacro{\xepi }{\bepi/\aepi*sqrt(\aepi^2-\c^2)}

    \draw[thick] (-\xepi,\c) -- (\xepi,\c);

    \draw[dash dot] (0,-\aepi-\eps) -- (0,\c+\eps);
    \draw[dash dot] (-\bepi-\eps,0) -- (\bepi+\eps,0);

    \draw(0,-\aepi-0.6) node[below] {Apex};

    \draw[thin] (-\xepi-\eps,\c) -- (-\bepi-2*\eps-2*\measx,\c);
    \draw[thin] (-\bepi-2*\eps,0) -- (-\bepi-2*\eps-2*\measx,0);
    \draw[thin] (-\eps,-\aendo) -- (-\bepi-2*\eps-\measx,-\aendo);
    \draw[thin] (-\eps,-\aepi) -- (-\bepi-2*\eps-2*\measx,-\aepi);
    \draw[thin,latex-latex] (-\bepi-\measx-\eps,0) -- node[above,rotate=90] {$a_\text{endo}$} ++(0,-\aendo);
    \draw[thin,latex-latex] (-\bepi-2*\measx-\eps,0) -- node[above,rotate=90] {$a_\text{epi}$} ++(0,-\aepi);
    \draw[thin,latex-latex] (-\bepi-2*\measx-\eps,0) -- node[above,rotate=90] {$c$} ++(0,\c);

    \draw[thin] (0,\c+2*\eps) -- (0,\c+2*\eps+2*\measx);
    \draw[thin] (\bendo,\eps) -- (\bendo,+\c+2*\eps+\measx);
    \draw[thin] (\bepi,\eps) -- (\bepi,+\c+2*\eps+2*\measx);
    \draw[thin,latex-latex] (\bendo,+\c+\eps+\measx) -- node[above] {$b_\text{endo}$} ++(-\bendo,0);
    \draw[thin,latex-latex] (\bepi,+\c+\eps+2*\measx) -- node[above] {$b_\text{epi}$} ++(-\bepi,0);

    \pgfmathsetmacro{\thisspringrealnextangle }{15-210/9}
    \foreach \angle in {15,\thisspringrealnextangle,...,-195}{
        \path (\angle:{\bepi} and {\aepi});\pgfgetlastxy{\XCoord}{\YCoord}

        \pgfmathsetmacro{\thisspringrealangle }{atan2(\YCoord/\aepi^2,\XCoord/\bepi^2)}
        \normalspring[\thisspringrealangle]{\angle:{\bepi} and {\aepi}}
    }

    \pgfmathsetmacro{\thisspringreals }{-\xepi+2*\xepi/3}
    \foreach \x in {-\xepi,\thisspringreals,...,\xepi}{
        \omnispring[90]{\x,\c}
    }

    \begin{scope}
        \clip(-\bepi-\eps,-\aepi-\eps) rectangle(\bepi+\eps,\c);
        \draw[thick] (0,0) ellipse ({\bendo} and {\aendo});
        \draw[thick] (0,0) ellipse ({\bepi} and {\aepi});
    \end{scope}

    \coordinate (pressurec) at (0,-0.5);
    \fill[white] (pressurec) circle(0.5);
    \node (pressure) at (pressurec) {$p$};
    \foreach \angle in {0,36,...,360}{
        \draw[-latex] (pressure) -- ++(\angle:0.5);
    }

    \pgfmathsetmacro{\xgammaendo}{\bendo/\aendo*sqrt(\aendo^2-(\c/2.0)^2)}
    \pgfmathsetmacro{\xgammaepi}{\bepi/\aepi*sqrt(\aepi^2-(\c/4.0)^2)}
    
    \draw (-\xepi,\c) ++(+0.3,0) -- ++(0.1, 0.6) node[above] {$\Gamma_0^\text{base}$};
    \draw (-\xgammaendo,\c/2.0) -- ++(+0.2, 0.1) node[right] {$\Gamma_0^\text{endo}$};
    \draw (-\xgammaepi,\c/4.0) -- ++(-0.6, 0.1) node[left] {$\Gamma_0^\text{epi}$};
\end{tikzpicture}
    \end{center}
    \caption{The prolate spheroid used as a model left ventricle for the simulation of G\&R
        of the myocardium. The base of the heart is fixed with omnidirectional springs and the
        pericardium is modeled with springs in reference normal direction to mimic the stiffness of the left atrium and the surrounding tissue. The ventricle is loaded
        with systolic pressure $p$.}
    \label{fig:geometry}       %
\end{figure}
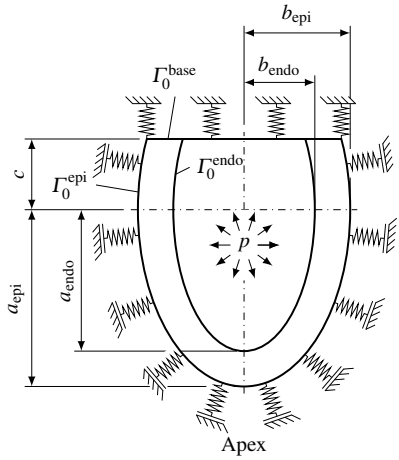

We implemented the above model in the in-house research code \cite{BACI} written in C++.
As often done in literature, we used the prolate spheroid shown in Figure~\ref{fig:geometry} as a simple model of a left ventricle. We assume that
cardiomyocytes and collagen fibers seek homeostasis in the systolic configuration. The dimensions of
the ventricle, therefore, represent a human LV in systole. We performed a mesh refinement study
and identified a mesh with 3970 quadratic tetrahedral elements to yield a good tradeoff between
computational efficiency and accuracy. The mesh was generated with Gmsh \citep{Geuzaine2009a}. We
use a Newton algorithm with backtracking to solve nonlinear systems of equations and a
conjugate gradient method for linear systems of equations \citep{trilinos-website}.

\begin{figure}
    \begin{center}
        \import{figures/}{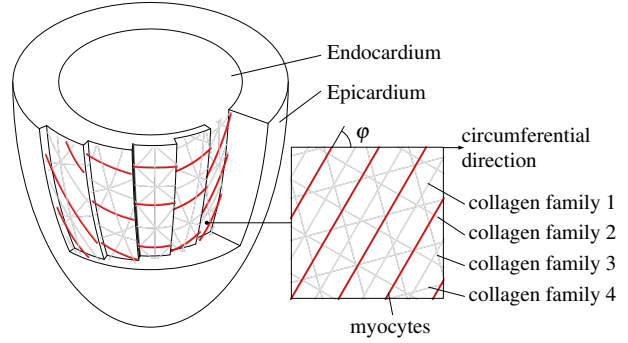}
    \end{center}
    \caption{The fiber distributions within the prolate spheroid. The helix angle $\varphi$ of Cardiomyocytes (red) varies from $+60^\circ$ at the endocardium
    to $-60^\circ$ at the epicardium. The four discrete collagen fiber families are aligned in myocyte fiber, cross-fiber and diagonal ($\pm 45^\circ$) directions.}
    \label{fig:fibers}       %
\end{figure}

The helix angle $\varphi$ of cardiomyocytes varies continuously from $+60^\circ$ at the endocardium to $-60^\circ$ at the
epicardium. We used the method described by \cite{Nagler2017a} to obtain the fiber directions of
cardiomyocytes at every point in the domain. Collagen fibers are represented with four discrete
fiber orientations that are aligned in the wall plane in myocyte fiber, cross-fiber, and diagonal ($\pm 45^\circ$)
direction (see Figure~\ref{fig:fibers}). The mass fractions for each collagen fiber direction are equal\-ly distributed.

The evolution equations (\ref{eqn:net_mass_production_rate}) and (\ref{eqn:remodeling_evolution_equation_oned})
are integrated with a forward Euler scheme using a step size that
is a twentieth of the minimum mean survival time ($T^i$ in equation (\ref{eqn:mass_degradation})).
\revb{
    We verified the integration of the local evolution equations with an implicit trapezoidal rule to exclude any numerical artifacts. All obtained results are identical within a small tolerance.
}

Figure\,\ref{fig:geometry} visualizes the geometry and the boundary conditions. To mimic the support of the left atrium, the base of the heart is supported by omnidirectional springs. The
epicardium is supported with springs in reference surface normal direction to capture the influence
of the pericardium and surrounding tissue \citep{pfaller2019a}. These springs are also prestressed during the prestress
algorithm as outlined in \cite{pfaller2019a}. On the endocardium, we apply a pressure to represent the
systolic pressure of the blood. The resulting virtual work formulation is
\begin{align}
    \begin{split}
        \delta W = &\int_{\mathcal{B}_0} \mat{P} : \delta \mat{F}\,  \dd V\\
         + &\int_{\Gamma_0^\text{base}} c_\text{base}\vct{u} \cdot \delta \vct{u} \, \dd \Gamma
        + \int_{\Gamma_0^\text{epi}} (c_\text{p} \vct{u} \cdot \vct{N}_0) \vct{N_0} \cdot \delta \vct{u} \, \dd \Gamma \\
        + &\int_{\Gamma_0^\text{endo}} p J \mat{F}^{-\T} \vct{N}_0 \cdot \delta \vct{u} \, \dd \Gamma= 0,
        \label{eqn:results_principle_of_virtual_work}
    \end{split}
\end{align}
where $\vct{N}_0$ is the reference outward surface normal, $\Gamma_0^\text{base}$ is the cut-off surface
of the ventricles, $\Gamma_0^\text{epi}$ and $\Gamma_0^\text{endo}$ are the epicardial and
endocardial surfaces, respectively.

The parameters used for the simulation can be found in Table~\ref{tab:parameters}. Material
parameters for constrained mixture type models of the myocardium have not yet been developed. Hence,
we adopted the material parameters and homeostatic stretches from \cite{braeu2016a} who developed a model of vascular G\&R.
The mean survival time of a sarcomere is estimated from the half-life time of the respective proteins
\citep{Willis2009a}. The stiffness of the pericardial boundary condition is taken from \cite{pfaller2019a}.
The initial mass fractions are roughly estimated from reported volume fractions in \cite{Holzapfel2009a}. The
dimensions of the used geometry were obtained from an MRI image of a 33-year-old
healthy female volunteer.

\begin{table}
    \caption{Simulation parameters used to study a hypertensive heart and reverse remodeling.
        The material parameters and homeostatic stretches are adopted from \cite{braeu2016a} who modeled
        vascular G\&R. The mean survival time of sarcomeres is estimated from \cite{Willis2009a}. The stiffness
        of the pericardial boundary conditions are taken from \cite{pfaller2019a} and initial mass
        fractions are estimated from volume fractions in healthy myocardium \citep{Holzapfel2009a}.}
        The classification of hypertension is taken from \cite{Chobanian2003a}.
    \label{tab:parameters}
    \begin{tabular}{llll}
        \hline\noalign{\smallskip}
        Name                                  & Parameter                       & Value                                  \\
        \noalign{\smallskip}\hline\noalign{\smallskip}
        \textit{Geometry}                                                                                                \\
        Major epicardial radius               & $a_\text{epi}$                  & $55\,\si{mm}$                          \\
        Minor epicardial radius               & $b_\text{epi}$                  & $30\,\si{mm}$                          \\
        Major endocardial radius              & $a_\text{endo}$                 & $46\,\si{mm}$                          \\
        Minor endocardial radius              & $b_\text{endo}$                 & $19\,\si{mm}$                          \\
        Truncation                            & $c$                             & $13\,\si{mm}$                          \\
        Epicardial fiber helix angle          & $\varphi_\text{epi}$            & $-60^\circ$                            \\
        Endocardial fiber helix angle         & $\varphi_\text{endo}$           & $60^\circ$                             \\
        \textit{Material parameters}                                                                                     \\
        Myocytes: Fung exponential parameters & $a^\text{m}$                    & $7.6\,\si{\nicefrac{J}{kg}}$           \\
        ~                                     & $b^\text{m}$                    & $11.4$                                 \\
        Myocytes: Active muscle tone          & $\sigma_{\text{act}}^\text{m}$  & $54\,\si{kPa}$                         \\
        ~                                     & $\lambda_{0}^\text{m}$          & $0.8$                                  \\
        ~                                     & $\lambda_{\text{max}}^\text{m}$ & $1.4$                                  \\
        Collagen: Fung exponential parameters & $a^\text{c}$                    & $568\,\si{\nicefrac{J}{kg}}$           \\
        ~                                     & $b^\text{c}$                    & $11.2$                                 \\
        Elastin: Neohookean parameter         & $a^\text{e}$                    & $72\,\si{\nicefrac{J}{kg}}$            \\
        Volumetric penalty                    & $\kappa$                        & $150\,\si{kPa}$                        \\
        \textit{Homeostatic stretches}                                                                                   \\
        Myocytes: Homeostatic stretch         & $\lambda_\text{h}^\text{m}$     & $1.1$                                  \\
        Collagen: Homeostatic stretch         & $\lambda_\text{h}^\text{c}$     & $1.062$                                \\
        \textit{Boundary conditions}                                                                                     \\
        Base spring stiffness                 & $c_\text{base}$                 & $2.0\,\si{\nicefrac{kPa}{mm}}$         \\
        Pericardial spring stiffness          & $c_\text{p}$                    & $0.2\,\si{\nicefrac{kPa}{mm}}$         \\
        Baseline systolic blood pressure      & $p$                             & $120\,\si{mmHg}$                       \\
        Stage 1 systolic blood pressure       & $p$                             & $140\,\si{mmHg}$                       \\
        Stage 2 systolic blood pressure       & $p$                             & $180\,\si{mmHg}$                       \\
        \textit{Initial conditions}                                                                                      \\
        Total inital reference mass density   & $\rho_0(s=0)$                   & $1050\,\si{\nicefrac{kg}{m^3}}$        \\
        Initial myocyte mass fraction         & $\xi^\text{m}(s=0)$             & $0.6$                                  \\
        Initial collagen mass fraction        & $\xi^\text{c}(s=0)$             & $0.1$                                  \\
        Initial elastin mass fraction         & $\xi^\text{e}(s=0)$             & $0.3$                                  \\
        Initial inelastic remodeling stretch  & $\lambda_\text{r}^i(s=0)$       & $\frac{1}{\lambda_\text{h}^i}$         \\
        \textit{G\&R parameters}                                                                                         \\
        Myocyte growth gain                   & $k^\text{m}$                    & $\nicefrac{0.1}{T^\text{m}}$           \\
        Myocyte sarcomere mean survival time  & $T^\text{m}$                    & $10\,\si{days}$                        \\
        Collagen growth gain                  & $k^\text{c}$                    & $\nicefrac{0.1}{T^\text{c}}$           \\
        Collagen mean survival time           & $T^\text{c}$                    & $15\,\si{days}$                        \\
        \textit{Stability criteria}                                                                                      \\
        Homeostasis threshold                 & $\epsilon_{\sigma,\text{h}}$    & $10^{-3}$                              \\
        \noalign{\smallskip}\hline
    \end{tabular}
\end{table}

\subsection{Prestress}
\label{sec:results:prestress}

\begin{figure*}
    \begin{center}
    \begin{tikzpicture}
        \begin{axis}[
                width=\linewidth/2.5,
                name=cell_mms_uncoupled_left,
                xmin=1, xmax=1000,
                ymin=1e-4, ymax=1e1,
                xlabel=\small{prestress iteration},
                ylabel=\small{$\max||\vct{u}||$ \(\left[\si{mm}\right]\)},
                ymode=log,
                grid=major,
                name=plot_displacement_norm,
            ]

            \addplot[mark=none, line width=1pt, color=dia3x3]
                table [x=timestep, y=max_nod_dis_norm, col sep=comma]
                {figures/data/prestress_over_time_6.csv};
            %\addlegendentry{$10^{-3}$}

            \addplot[mark=none, line width=1pt, color=dia3x2, dashed ]
                table [x=timestep, y=max_nod_dis_norm, col sep=comma]
                {figures/data/prestress_over_time_5.csv};
            %\addlegendentry{$10^{-2}$}

            \addplot[mark=none, line width=1pt, color=dia3x1, dotted ]
                table [x=timestep, y=max_nod_dis_norm, col sep=comma]
                {figures/data/prestress_over_time_4.csv};
            %\addlegendentry{$10^{-1}$}
        \end{axis}

        \pgfplotsset{minor grid style={gray}}
        \pgfplotsset{major grid style={gray!50}}
        \begin{axis}[
                width=\linewidth/2.5,
                name=cell_mms_uncoupled_left,
                xmin=1, xmax=1000,
                ymin=1e-5, ymax=1e-1,
                xlabel=\small{prestress iteration},
                ylabel=\small{$\max|\frac{\sigma^\text{m}-\sigma_\text{h}^\text{m}}{\sigma_\text{h}^\text{m}}|$},
                ymode=log,
                grid=major,
                name=plot_myocyte_stress,
                at={($(plot_displacement_norm.east)+(2cm,0)$)},
                anchor=west,
                legend cell align=left,
                legend pos=outer north east
            ]

            \addplot[mark=none, line width=1pt, color=dia3x3 ]
                table [x=timestep, y=fiber_stress_error, col sep=comma]
                {figures/data/prestress_over_time_6.csv}; \label{pgfplots:prestress_plot_e6};
            \addlegendentry{$\varepsilon_\text{pre} = 10^{-3}\,\si{mm}$}

            \addplot[mark=none, line width=1pt, color=dia3x2, dashed]
                table [x=timestep, y=fiber_stress_error, col sep=comma]
                {figures/data/prestress_over_time_5.csv}; \label{pgfplots:prestress_plot_e5};
            \addlegendentry{$\varepsilon_\text{pre} = 10^{-2}\,\si{mm}$}

            \addplot[mark=none, line width=1pt, color=dia3x1, dotted]
                table [x=timestep, y=fiber_stress_error, col sep=comma]
                {figures/data/prestress_over_time_4.csv}; \label{pgfplots:prestress_plot_e4};
            \addlegendentry{$\varepsilon_\text{pre} = 10^{-1}\,\si{mm}$}
        \end{axis}

        \node[below = 1.0cm of plot_displacement_norm.south] {(a)};
        \node[below = 1.0cm of plot_myocyte_stress.south] {(b)};
    \end{tikzpicture}
\end{center}
    \caption{Convergence of the prestress algorithm. After prestretching, G\&R is activated with
    unchanged boundary conditions. (a) Maximum Euclidean norm of the nodal displacements between the
        reference configuration and the prestressed configuration for different convergence thresholds. After the
        subsequent G\&R phase, the maximum Euclidean norm of the nodal displacements is around half of an order of magnitude higher than
        the chosen prestress convergence thresholds. (b) Maximum relative deviation of the myocyte fiber Cauchy
        stress from the homeostatic stress for different convergence thresholds.}
    \label{fig:prestress_time_curves}
\end{figure*}
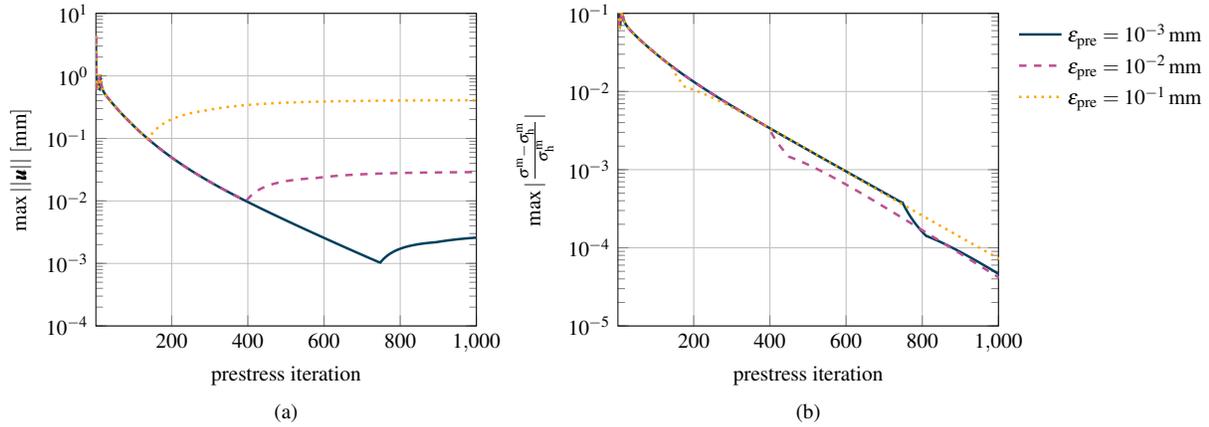

\begin{figure*}
    \begin{center}
    \begin{tikzpicture}
        \import{figures/data/}{pr_eigenvalue1.tex}
        \begin{axis}[
                hide axis,
                width=\linewidth/3,
                scale only axis,
                enlargelimits=false,
                xmin=-\thisimagehalfwidth,
                xmax=\thisimagehalfwidth,
                ymin=-\thisimagehalfheight,
                ymax=\thisimagehalfheight,
                axis equal=true,
                name=plot2,
            ]
            \addplot[thick,blue] graphics[xmin=-\thisimagehalfwidth,ymin=-\thisimagehalfheight,xmax=\thisimagehalfwidth,ymax=\thisimagehalfheight] {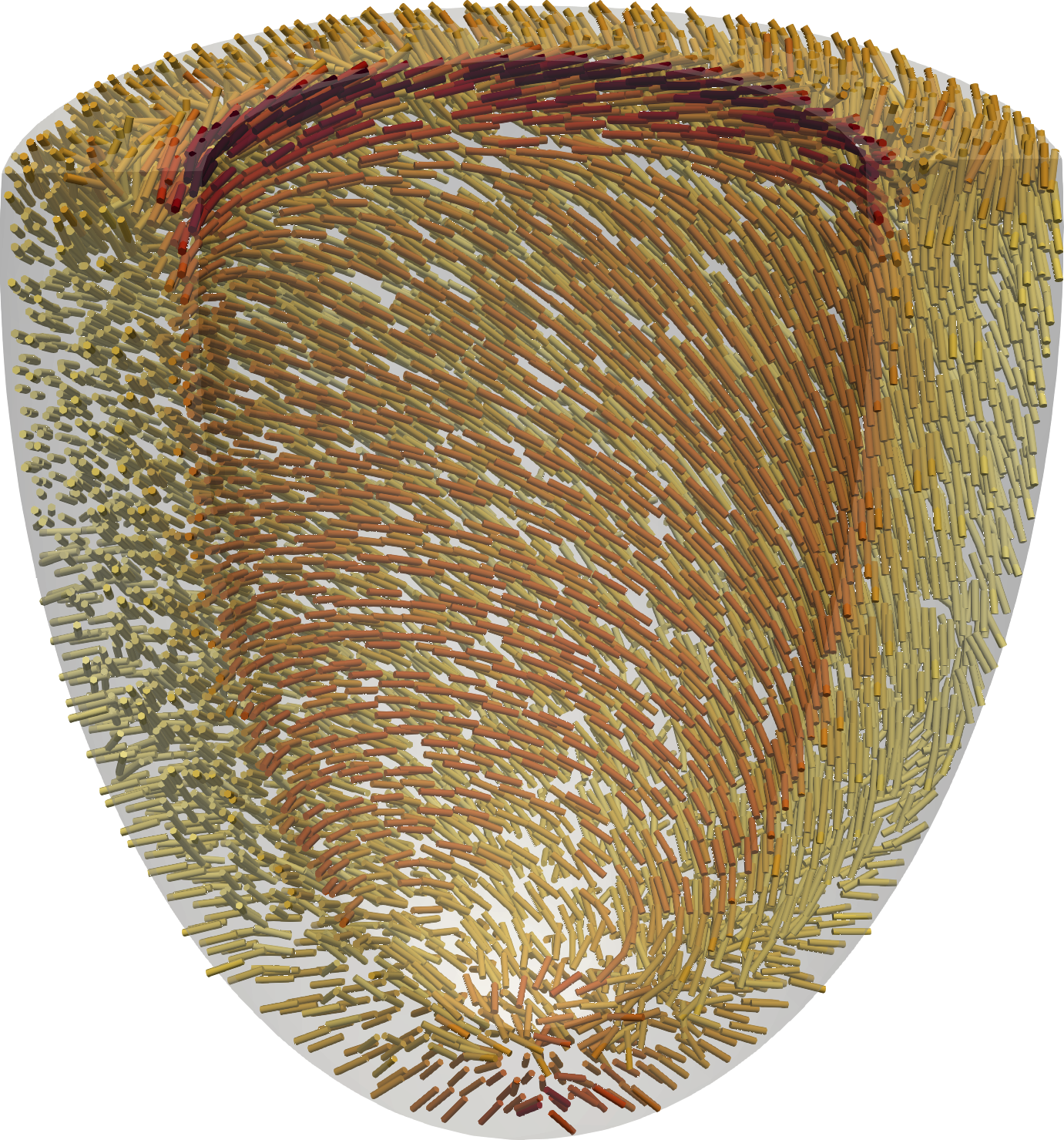};
        \end{axis}

        \import{figures/data/}{pr_eigenvalue3.tex}
        \begin{axis}[
                hide axis,
                width=\linewidth/3,
                scale only axis,
                enlargelimits=false,
                xmin=-\thisimagehalfwidth,
                xmax=\thisimagehalfwidth,
                ymin=-\thisimagehalfheight,
                ymax=\thisimagehalfheight,
                axis equal=true,
                name=plot3,
                at={(plot2.north east)},
                anchor=north west
            ]
            \addplot[thick,blue] graphics[xmin=-\thisimagehalfwidth,ymin=-\thisimagehalfheight,xmax=\thisimagehalfwidth,ymax=\thisimagehalfheight] {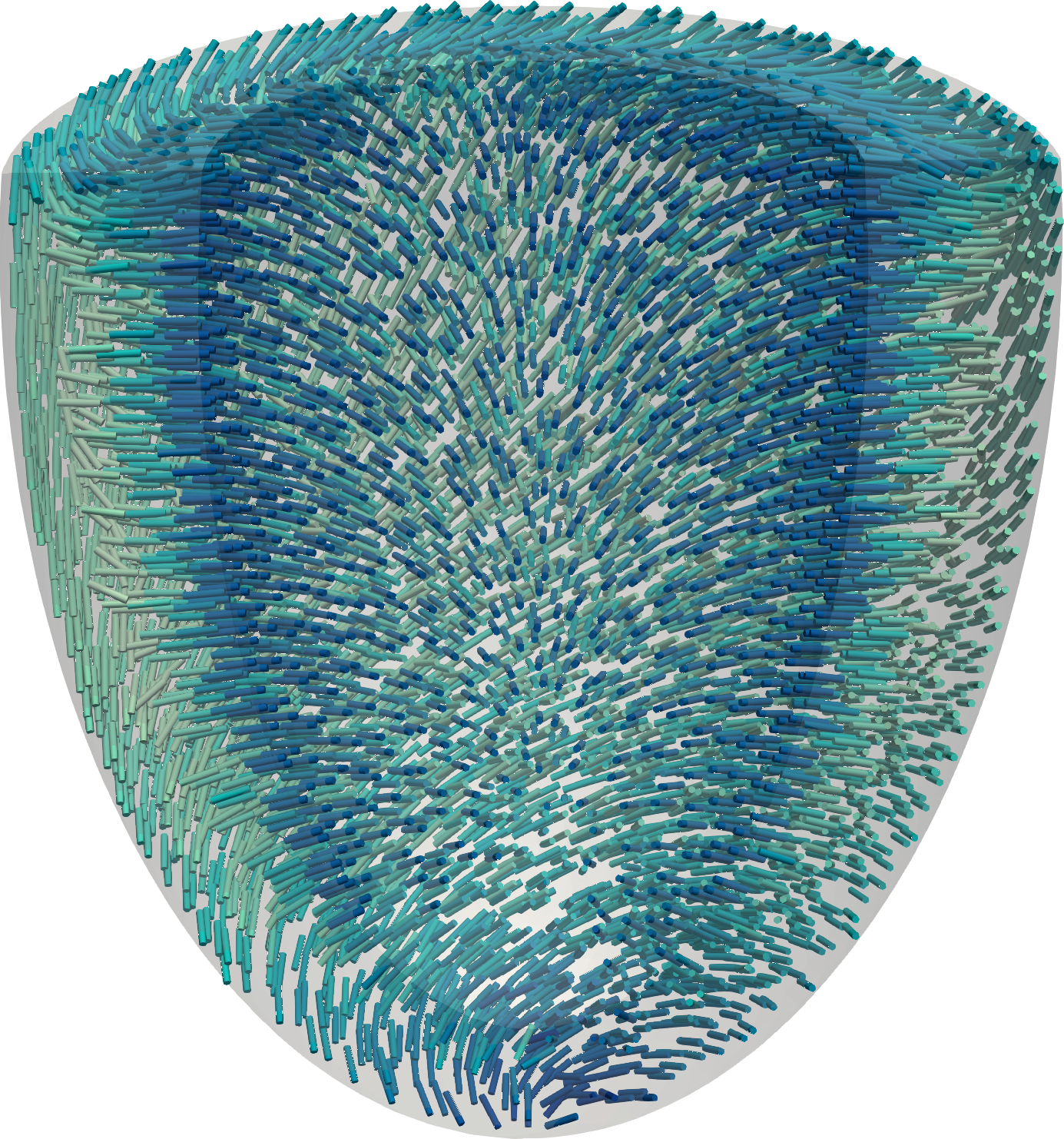};
        \end{axis}
        \begin{axis}[
                hide axis,
                colormap name=thisplotcolormap,
                colorbar,
                colorbar style={%
                        ylabel=Principal stretch,
                    },
                point meta max=\eigenvaluerangemax,
                point meta min=\eigenvaluerangemin,
                name=plotcolorbar,
                at={(plot3.north east)},
                anchor=north west
            ]
        \end{axis}

        \node[below = 0.5cm of plot2.south] {(a)};
        \node[below = 0.5cm of plot3.south] {(b)};

    \end{tikzpicture}
\end{center}
    \caption{Visualization of the elastin prestretch. (a) Principal direction and principal stretch of the largest Elastin
    expansion (first principal stretch) and
    (b) for the largest compression (third principal stretch) at every integration point, respectively.}
    \label{fig:prestress_3d}
\end{figure*}

In the following, we analyze the prestress algorithm presented in section~\ref{sec:modeling:prestress}.
The goal is to obtain the prestretch tensor of elastin at each Gauss point such that the reference configuration
is in equilibrium with the systolic pressure $p=120\,\si{mmHg}$ applied at the endocardium. We linearly ramp
up the pressure and the homeostatic stretch of the G\&R constituents in the first 10 timesteps. We then run
our prestressing algorithm until the maximum Euclidean norm of the nodal displacements falls below $\varepsilon_\text{pre}$. Finally,
we terminate the prestressing algorithm and activate our G\&R model to equilibrate any residual deviations
from the homeostatic state.

We investigate 3 different prestress convergence thresholds $\varepsilon_\text{pre} \in \left\{10^{-1}, 10^{-2}, 10^{-3}\right\}\,\si{mm}$.
Since the displacements of the prestressed configuration are not exactly zero, the constituents are
not in perfect but approximate homeostasis. We thus also investigate in this section how the myocardium
approaches a stable homeostatic configuration after convergence of the prestress algorithm.

Figure~\ref{fig:prestress_time_curves} (a) shows the maximum Euclidean norm of the nodal displacements over
the prestress iterations for each simulation and (b) shows the respective relative deviation from homeostasis
for the cardiomyocytes.

After $136$ timesteps, the maximum nodal displacement falls below the first threshold
of $\varepsilon_\text{pre} = 0.1\,\si{mm}$. The remaining displacements cause a maximum relative deviation
from homeostasis of the cardiomyocytes of $2\,\%$. When activating
G\&R at this stage, G\&R converges towards a limit configuration with a final maximum nodal
displacement of $0.4\,\si{mm}$.

If we choose the threshold $\varepsilon_\text{pre} = 0.01\,\si{mm}$ for the prestress algorithm, $393$ iterations are needed to fulfill
the criterion. The deviation from homeostasis for cardiomyocytes is $0.4\,\si{\%}$.
The maximum nodal displacement after converged G\&R is $0.03\,\si{mm}$.

The lowest investigated threshold $\varepsilon_\text{pre} = 0.001\,\si{mm}$ for the prestress algorithm is achieved after $745$ iterations with $0.04\,\si{\%}$ deviation from homeostasis. The maximum nodal displacement after subsequent G\&R is $0.003\,\si{mm}$.

Figure~\ref{fig:prestress_time_curves} (b) depicts the maximum deviation from homeostasis in the myocardium during prestressing and the subsequent G\&R phase. The convergence behavior towards me\-chano\-bio\-logical equilibrium, characterized by the slope of the three curves, is similar in all three cases during prestressing and the subsequent G\&R phase.

In the following, we analyze the prestretch of elastin. We analyze the case with $\varepsilon_\text{pre}=0.01\,\si{mm}$
after convergence of the prestress algorithm.
Figure~\ref{fig:prestress_3d} (a) depicts the direction of the largest principal elastin prestretch (largest elastin expansion)
and (b) for the smallest principal stretch (largest elastin compression) at every 
integration point in the myocardium, respectively.

The largest principal stretch occurs at the basal plane at the endocardium, with a principal stretch
of around $1.5$.
A slightly lower principal stretch occurs in the apical region. %
Generally, the principal stretches are larger at the endocardium and smaller at the epicardium.
At the en\-do\-car\-di\-um, elastin is under expansion in the plane of cardiomyocytes and collagen fibers,
and under compression perpendicular to the plane (to bear the load of the blood pressure). On the
epicardium, elastin is under expansion perpendicular to the fiber plane, where the springs of the
pericardial boundary condition pull on the contracted ventricle (see Figure~\ref{fig:prestress_3d}).

\subsection{Mechanobiological stability}
\label{sec:results:stability}

\begin{figure}
    % !TEX root = ../paper.tex
\pgfplotsset{
    discard if not A/.style 2 args={
        x filter/.append code={
            \edef\tempa{\thisrow{#1}}
            \edef\tempaa{#2}
            \ifx\tempa\tempaa
            \else
                \def\pgfmathresult{}
            \fi
        }
    }}
\begin{tikzpicture}
    \begin{axis}[
            width=\linewidth,
            name=cell_mms_uncoupled_left,
            xmin=130.0, xmax=240,
            ymin=0.05, ymax=0.2,
            xlabel=\small{Systolic pressure \(p \left[\si{mmHg}\right]\)},
            ylabel=\small{Growth constant \(k^i \cdot T^i \left[\si{-}\right]\)},
            legend style={font=\footnotesize, draw=black, fill=white},
            legend pos=south east
        ]
        
        \addplot[only marks,mark=o, color=dia3x3, line width=1.5pt, mark size=2pt]
            table [x=pressure, y=growth_gain, col sep=comma, discard if not A={homeostasis}{True}]
            {figures/data/stability_map.csv};
        \addlegendentry{stable};
        
        \addplot[only marks,mark=x, color=dia3x2, line width=1.5pt, mark size=3pt]
            table [x=pressure, y=growth_gain, col sep=comma, discard if not A={stable}{False}]
            {figures/data/stability_map.csv};
        \addlegendentry{unstable};
        
        \draw[dia3x3,line width=1pt] (axis cs:140,0.1) -- (axis cs:135,0.11);
        \draw[dia3x3,fill=white,line width=1pt] (axis cs:135,0.11) circle(0.18cm) node {$1$};
        \addplot[only marks,mark=*, color=dia3x3, line width=1.5pt, mark size=2pt, mark options={fill=white}] coordinates {(140,0.1)};

        \draw[dia3x2,line width=1pt] (axis cs:180,0.1) -- (axis cs:180,0.11);
        \draw[dia3x2,fill=white,line width=1pt] (axis cs:180,0.11) circle(0.18cm) node {$2$};
        
        % undetermined simulations (either unstable but homeostasis or stable but not in homeostasis) -> requires further study
        \addplot[only marks,mark=diamond, color=dia3x1]
            table [x=pressure, y=growth_gain, col sep=comma, discard if not A={homeostasis}{False}, discard if not A={stable}{True}]
            {figures/data/stability_map.csv};
        
        \addplot[only marks,mark=diamond, color=dia3x1]
            table [x=pressure, y=growth_gain, col sep=comma, discard if not A={homeostasis}{True}, discard if not A={stable}{False}]
            {figures/data/stability_map.csv};
    \end{axis}
\end{tikzpicture}
    \caption{Mechanobiological stability map of heart growth depending on the systolic pressure and the growth
        constant $k^\text{i}\cdot T^i$. Blue circles mark simulations that result in a stable end configuration,
        whereas simulations with purple crosses mark cases with unbounded growth. \revc{The cases (1) and (2) are shown in more detail in Figure~\ref{fig:gr_time_curves}.}}
    \label{fig:stability_map}
\end{figure}
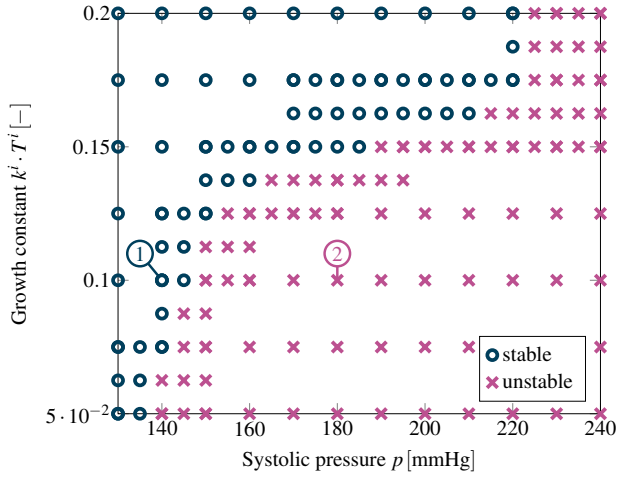

\begin{figure*}
    \import{figures/}{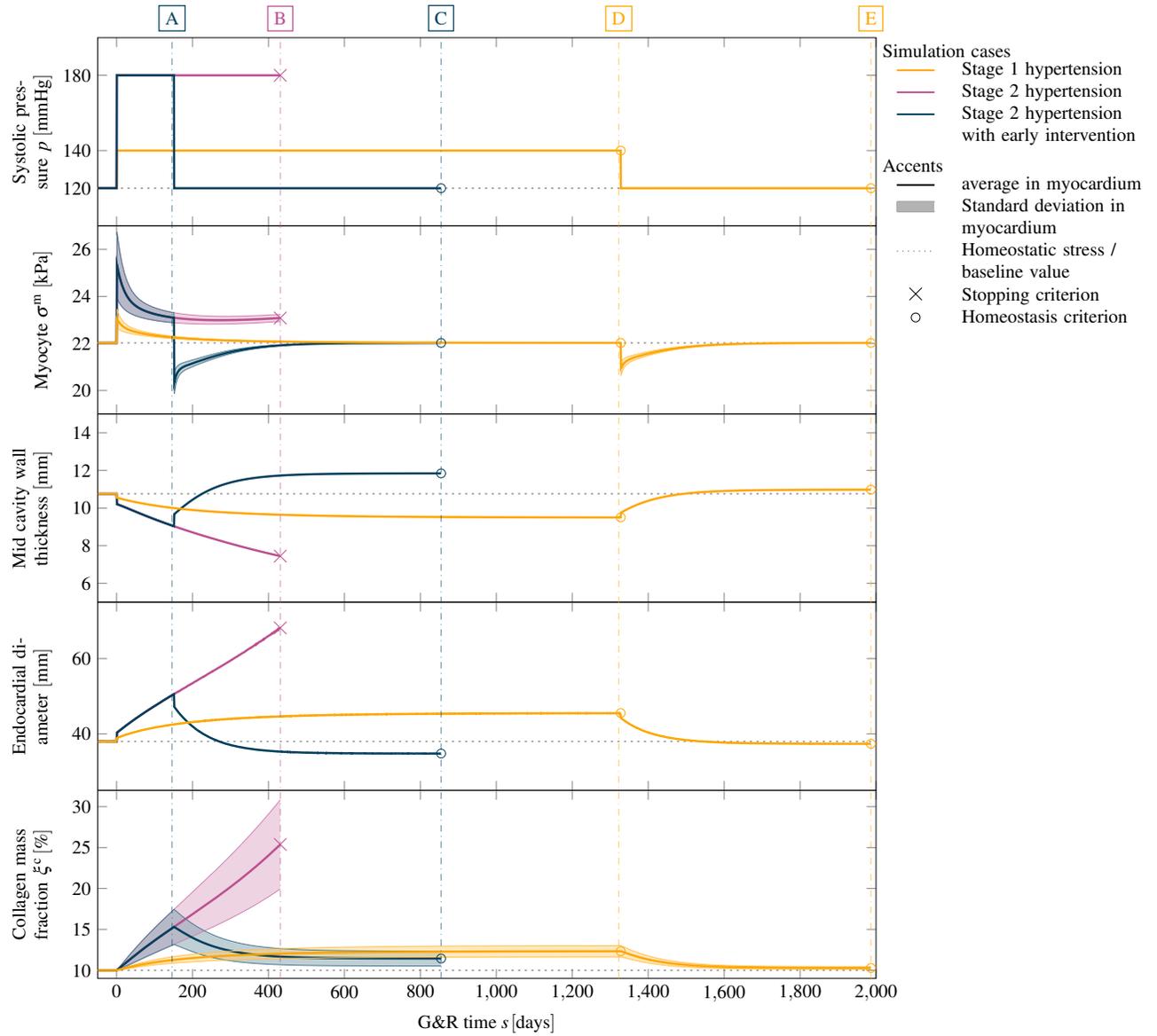}
    \caption{Stage 1 and stage 2 hypertension of a left ventricle with a subsequent pressure
    normalization and resulting reversal of G\&R. For each case, we plot (from top to bottom) the applied systolic
    pressure, the resulting Cauchy stress of cardiomyocytes with standard deviation, the mid cavity wall thickness, the endocardial diameter
    and the collagen mass fraction with the standard deviation in the myocardium. Stage 1 hypertension resulted in stable G\&R. After a normalization of the systolic pressure, all quantities return back close to their
    pre-G\&R reference values. Both configurations [D] and [E] are depicted in Figure~\ref{fig:case140finalimage}.
    Stage 2 hypertension results in unstable G\&R.
    After the deviation from homeostasis of all constituents consistently increased, the simulation is stopped. This configuration [B] is depicted
    in Figure~\ref{fig:case180finalimage}. If hypertension is treated after $150\,\si{days}$ (configuration [A])
    the reverse G\&R phase results in a stable configuration [C]. Mid cavity wall thickness, endocardial
    diameter and collagen mass fraction differ from the pre-G\&R reference values. Both configurations [A] and [C] are depicted in Figure~\ref{fig:case180_early_reversal_3d}.
    \revb{In the supplementary material, we also show the evolution of the Cauchy fiber stress of each collagen fiber family and the
    evolution of the mass fraction of cardiomyocytes and all collagen fiber families individually.}}
    \label{fig:gr_time_curves}
\end{figure*}

\begin{figure*}
    \begin{center}
    \begin{tikzpicture}
        \import{figures/data/}{case140-1_gr_last.tex}
        \begin{axis}[
                hide axis,
                width=6.5cm,
                scale only axis,
                enlargelimits=false,
                xmin=-569.5,
                xmax=569.5,
                ymin=-606,
                ymax=606,
                axis equal=true,
                name=plot1,
            ]
            \addplot[thick,blue] graphics[xmin=-\thisimagehalfwidth,ymin=-\thisimagehalfheight,xmax=\thisimagehalfwidth,ymax=\thisimagehalfheight] {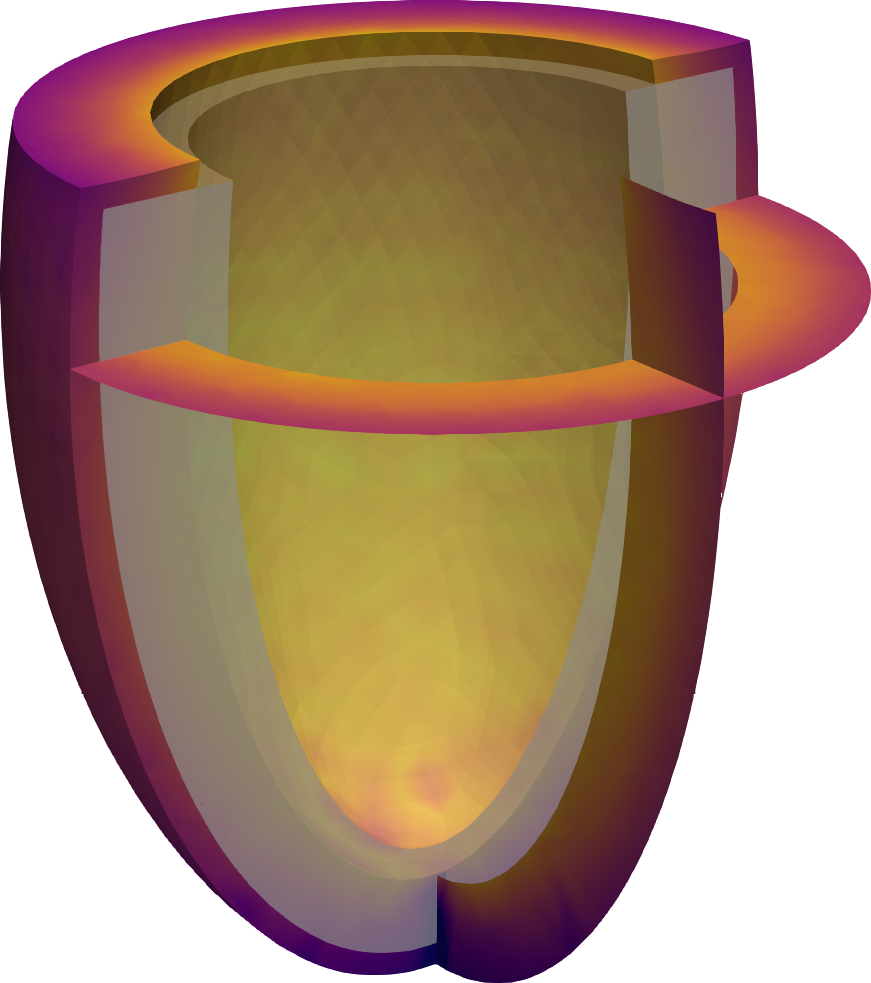};
        \end{axis}

        \import{figures/data/}{case140-2_gr_last.tex}
        \begin{axis}[
                hide axis,
                width=6.5cm,
                scale only axis,
                enlargelimits=false,
                xmin=-569.5,
                xmax=569.5,
                ymin=-606,
                ymax=606,
                axis equal=true,
                name=plot2,
                at={(plot1.north east)},
                anchor=north west
            ]
            \addplot[thick,blue] graphics[xmin=-\thisimagehalfwidth,ymin=-\thisimagehalfheight,xmax=\thisimagehalfwidth,ymax=\thisimagehalfheight] {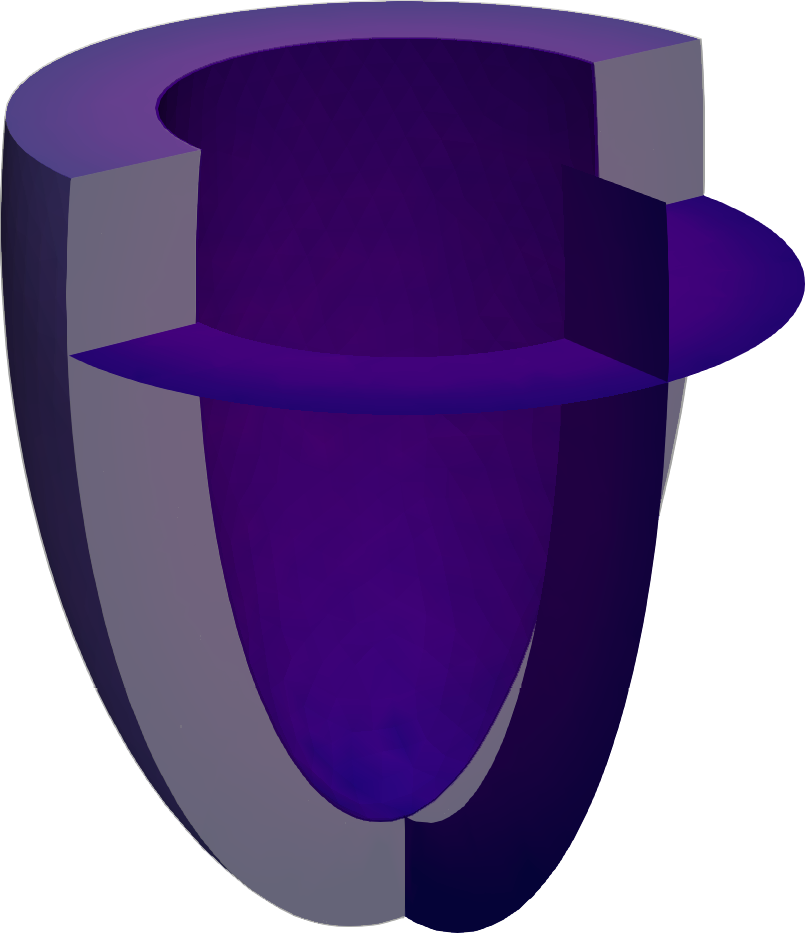};
        \end{axis}
        \begin{axis}[
                hide axis,
                colormap name=thisplotcolormap,
                colorbar,
                colorbar style={%
                    yticklabel={\pgfmathprintnumber\tick\,\%},
                    ylabel=Local mass increase,
                },
                point meta max=\volumechangemax,
                point meta min=\volumechangemin,
                name=plotcolorbar,
                at={(plot2.north east)},
                anchor=north west
            ]
        \end{axis}

        \node[color=dia3x1, draw, above = 0cm of plot1.south, anchor=center] {D};
        \node[color=dia3x1, draw, above = 0cm of plot2.south, anchor=center] {E};

    \end{tikzpicture}
\end{center}
    \caption{The final configuration of each G\&R phase of the simulated stage 1 hypertension. The reference configuration is depicted in gray. Configuration [D] has recovered homeostasis
        for an increased pressure of $p=140\,\si{mmHg}$ (stage 1 hypertension). The cavity volume of the ventricle increased and a slightly
        more spherical shape is observed. Configuration [E] has recovered homeostasis after returning to
        baseline pressure $p=120\,\si{mmHg}$. An almost complete reversal of the growth is observed
        resulting in only a small deviation from the reference configuration.}
    \label{fig:case140finalimage}
\end{figure*}

\begin{figure}
    \begin{center}
    \begin{tikzpicture}
        \import{figures/data/}{case180-1_gr_last.tex}
        \begin{axis}[
                hide axis,
                width=6.5cm,
                scale only axis,
                enlargelimits=false,
                xmin=-569.5,
                xmax=569.5,
                ymin=-606,
                ymax=606,
                axis equal=true,
                name=plot2,
            ]
            \addplot[thick,blue] graphics[xmin=-\thisimagehalfwidth,ymin=-\thisimagehalfheight,xmax=\thisimagehalfwidth,ymax=\thisimagehalfheight] {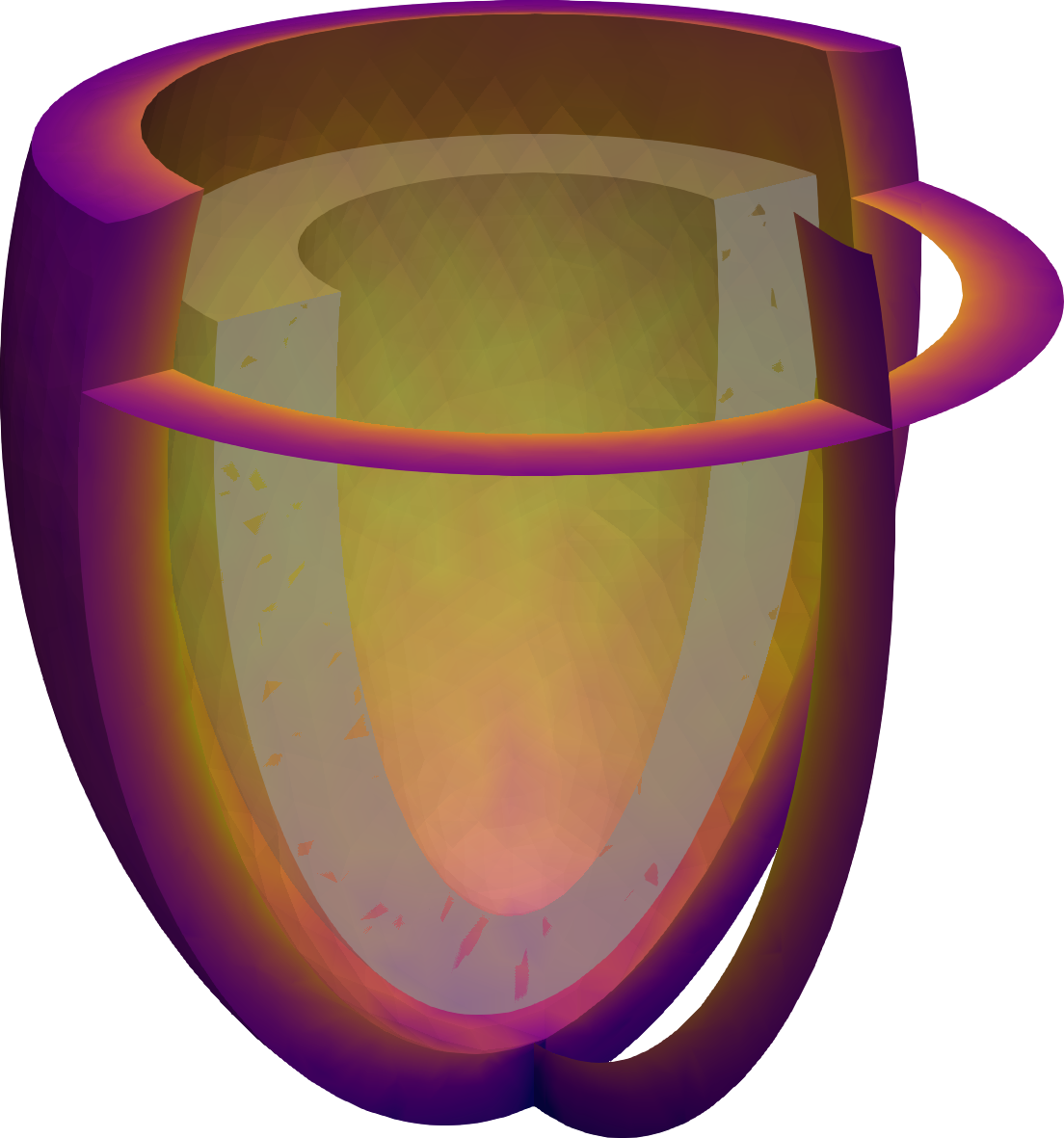};
            
            % TODO: adapt figure with current "end" of the simulation
        \end{axis}
        \begin{axis}[
                hide axis,
                colormap name=thisplotcolormap,
                colorbar,
                colorbar style={%
                    yticklabel={\pgfmathprintnumber\tick\,\%},
                    ylabel=Local mass increase,
                },
                point meta max=\volumechangemax,
                point meta min=\volumechangemin,
                xticklabel = \pgfmathprintnumber\tick\%,
                name=plotcolorbar,
                at={($(plot2.east)+(-1.2cm,0)$)},
                anchor=west
            ]
        \end{axis}
        
        \node[color=dia3x2, draw, below = 0cm of plot2.south] {B};
        
        %\node[left = 0cm of plot2.west, anchor=center, rotate=90] {stage 2 hypertension};

    \end{tikzpicture}
\end{center}
    \caption{Configuration [B] after stopping unstable G\&R for a systolic pressure of $p=180\,\si{mmHg}$
    (stage 2 hypertension). The cavity size increased and the shape renders spherically.}
    \label{fig:case180finalimage}
\end{figure}

In the following, we examine mechanobiological stability of the model of cardiac G\&R described in Section~\ref{sec:modeling} depending on external loading and mechanobiological key parameters.

To this end, we performed a series of G\&R simulations all starting from a prestressed configuration at $p=120\,\si{mmHg}$ obtained with the procedure described in
Section~\ref{sec:modeling:prestress} (using a prestress convergence threshold of
$\varepsilon_\text{pre}=0.01\,\si{mm}$ with a subsequent activation of G\&R equations for 100 timesteps
to bring the constituents close to homeostasis). Starting from such a configuration, we increased the systolic pressure by a Heaviside step function to a new value $p^+$. We examined the response of the LV to this perturbation in a parameter study. To this end,
we varied the growth gain parameters of all remodeling
constituents and the systolic blood pressure. The blood pressure was varied between 
$130\,\si{mmHg}$ and $240\,\si{mmHg}$ such that stage 1 and stage 2 hypertension are covered \citep{Chobanian2003a}.
In the first step, the parameter region
of interest, that is, the two-dimensional parameter space $\{(k^i,p^+):~ \nicefrac{0.05}{T^i} \le k^i \le \nicefrac{0.2}{T^i};~130\,\si{mmHg} \le p^+ \le 240\,\si{mmHg}\}$, is sampled with a spacing of $\Delta k^i = \nicefrac{0.025}{T^i}$, $\Delta p^+ = 10\,\si{mmHg}$ using a coarse mesh with $2118$ quadratic tetrahedral elements. In this first step it is observed that in one part of this parameter space one observed mechanobiologically unstable G\&R, in the other one, however, mechanobiologically stable G\&R. To determine the
subregion between stability and instability more exactly, 
the parameter region around the border between stable and unstable G\&R is resampled in a second step with a finer spacing in the parameter space. In a last step, the refined border is verified with simulations using the fine mesh with 3970 elements identified
in the mesh convergence study.

A simulation is \emph{mechanobiologically stable} if the systems returns to a (new) mechanobiological equilibrium \citep{Cyron2014a} after the perturbation of the external loading. In mechanobiological equilibrium, all constituents are in their preferred stress state, i.e.
\begin{align*}
    \lim_{s \rightarrow{} \infty} \frac{|\sigma^i-\sigma_\text{h}^i|}{\sigma_\text{h}^i} = 0 \quad \forall \quad i.
\end{align*}
Otherwise, a simulation exhibits unstable runaway growth and will not converge to a stable homeostatic configuration. We approximate this classification in the context of a standard forward simulation. A simulation is stopped, once the criterion
\begin{align*}
    \max \frac{|\sigma^i-\sigma_\text{h}^i|}{\sigma_\text{h}^i} \le \epsilon_{\sigma,\text{h}} \quad \forall \quad i%
\end{align*}
is fulfilled (\emph{mechanobiological stability}). If the deviation from homeostasis increases for all constituents, we classify the simulation as  \emph{mechanobiologically unstable}:
\begin{align*}
    \frac{\dd}{\dd s}\frac{|\sigma^i-\sigma_\text{h}^i|}{\sigma_\text{h}^i} > 0 \quad \forall \quad i.
\end{align*}

We ran all samples long beyond the point of the stability or instability stopping criteria as given above. None of the samples changed its classification back to instability once the instability criterion was fulfilled or vice versa within the considered timeframe.

Figure~\ref{fig:stability_map} visualizes the classification of each sample. In case of a
slightly elevated pressure ($p^+ \le 135\,\si{mmHg}$ ) G\&R is stable for all studied growth constants. In contrast,
a highly elevated pressure ($p^+ \ge 225\,\si{mmHg}$) always results
in unstable growth. In between, stability depends on the growth constant, i.\,e. how fast the
heart can adapt to the change in LV systolic pressure. The higher the
growth constant $k^i \cdot T^i$, the higher the range of increased pressure that yields a stable configuration.

\reva{We have shown the influence of the systolic pressure and the growth constant on mechanobiological stability.
In the supplementary materials, we show that the stiffness of the spring boundary conditions also influences mechanobiological
stability. These results show that
the higher the spring stiffness, the smaller is the deviation of the fiber Cauchy stress from homeostasis. The stiffer boundary conditions can bear more of the
load. As a consequence, the results of G\&R are less pronounced (smaller endocardial diameter and thinner mid-cavity wall thickness). The mechanobiological stability
region is larger if the spring stiffness is higher.}

In Figure~\ref{fig:gr_time_curves}, we focus on one sample of each region, stable and unstable. For both cases, we assume a
growth constant of $k^\text{i} \cdot T^i=0.1$, i.\,e. the same material properties. Both cases
differ just by the applied LV systolic pressure increase starting from a baseline homeostatic
pressure of $p=120\,\si{mmHg}$. For the stable case, we assume a systolic blood pressure of $p^+=140\,\si{mmHg}$ which would
be classified as a stage 1 hypertension \citep{Chobanian2003a}. For the unstable case, we assume an
increase of $50\,\si{\%}$ ($p^+=180\,\si{mmHg}$) which would be classified as stage 2 hypertension.

Figure~\ref{fig:gr_time_curves} shows for both cases the applied systolic pressure, the mean of the Cauchy fiber
stress of cardiomyocytes at all Gauss points with standard deviation, the relative change of the wall thickness
in the mid-cavity, the endocardial diameter over time, and the collagen mass fraction. In this
section, we discuss stage 1 hypertension (yellow curve) and stage 2 hypertension (purple curve)
in the time interval of the increased pressure ($s \in [0, 1325]\,\si{days}$).

The sudden increase in left ventricular systolic blood pressure results in an elastic elongation of the
fibers and therefore in increased stress. 
In case of stage 1 hypertension, the cardiomyocyte fiber stress increases from the homeostatic value of $22\,\si{kPa}$ to $22.9 \pm 0.3\,\si{kPa}$ (mean $\pm$ standard deviation).
The discontinuity in the wall thickness and the endocardial diameter stems from the elastic elongation
of the ventricle due to the sudden pressure increase. The cardiomyocytes and respectively the
collagen fibers \revb{(shown in the supplementary material)} are not in their homeostatic state after the change in external loading and the G\&R
evolution equations stimulate G\&R of all fiber constituents. The stress \revb{of} cardiomyocytes \revb{and collagen fibers} decay exponentially toward
their homeostatic stress resulting in a stable limit configuration. In this configuration, the
mid-cavity wall thickness decreased to $9.5\,\si{mm}$ ($-12\,\%$ compared to the reference configuration). The endocardial
diameter increased to a value of $45\,\si{mm}$ ($+20\,\si{\%}$). Collagen fiber mass fraction
in the myocardium also increased from initially $10\,\%$ to $12 \pm 1\,\%$ (mean $\pm$ standard deviation). \revb{In the supplementary material, we also
show the evolution of the mass fractions of cardiomyocytes and all collagen fiber families individually.} The deposition of collagen fibers
is inhomogeneous within the myocardium (not shown) where the maximum is at the endocardium and the minimum at the epicardium. Figure~\ref{fig:case140finalimage} (a)
depicts the configuration with the local mass change encoded in color. The reference configuration
is also drawn in light gray. Mass is mainly produced in the endocardium, where mass increased by about $10\,\si{\%}$.

The elastic deformation due to stage 2 hypertension results in a sudden increase in the
Cauchy stress of cardiomyocytes to $25.0 \pm 1.2\,\si{kPa}$. The elastic change of the mid-cavity wall thickness
and the endocardial diameter is larger than in stage 1 hypertension. First, the activation of the G\&R equations
lets the Cauchy stress decay towards the homeostatic stress. However, Cauchy stress plateaus
above the homeostatic stress with a subsequent increase. The simulation is stopped when the deviation from homeostasis increases for all constituents. At this point, the mid cavity wall thickness continuously decreased to $7.4\,\si{mm}$ ($-30\,\%$ compared to the
reference wall thickness)
and the endocardial diameter increased to $68\,\si{mm}$ ($+79\,\si{\%}$). Collagen mass is also excessively deposited resulting in a mass fraction of $25 \pm 5\,\%$ in the final configuration. The
final (unstable) configuration is shown in Figure~\ref{fig:case180finalimage}. Locally at the endocardium, the mass has increased by about $92\,\si{\%}$ compared to the reference configuration. The mass increase is lower at the epicardium.

\subsection{Reversal}

In the following section, we study to what extent G\&R can be reversed when the pressure returns to its baseline value.
Physiologically, this corresponds to an intervention that treats hypertension by lowering blood pressure, e\,g., through ACE inhibitors.

In this section, we focus on the intervals with pressures restored to baseline values in Figure~\ref{fig:gr_time_curves}  (i.e., for stage 1 hypertension the interval $s > 1325\,\si{days}$, i.e., after [D], and for stage 2 hypertension (with early intervention) the interval $s > 150\,\si{days}$, i.e., after [A]).
The starting point for the case stage 1 hypertension is a stable grown configuration as discussed
in Section~\ref{sec:results:stability} (yellow, configuration [D]). The elastic response of returning 
systolic pressure to the physiological level is an immediately decreased cardiomyocyte fiber stress ($21.0 \pm 0.3\,\si{kPa}$), a slight
increase in wall thickness and a slight decrease in endocardial diameter. The activation of the G\&R
lets the cardiomyocyte fiber stress return exponentially to its homeostatic value. The wall thickness
of the mid cavity increases slightly above the reference wall thickness ($+2.0\,\si{\%}$) and to
an almost identical endocardial diameter ($-1.6\,\si{\%}$) as in the reference configuration. Collagen mass fraction also returned to the baseline value of $10\,\%$. The final configuration [E] is
depicted in Figure~\ref{fig:case140finalimage}. It shows only small differences to the reference configuration depicted in light gray.

\begin{figure*}
    \begin{center}
    \begin{tikzpicture}
        \import{figures/data/}{case180b-1_gr_last.tex}
        \begin{axis}[
                hide axis,
                width=6.5cm,
                scale only axis,
                enlargelimits=false,
                xmin=-569.5,
                xmax=569.5,
                ymin=-606,
                ymax=606,
                axis equal=true,
                name=plot1,
            ]
            \addplot[thick,blue] graphics[xmin=-\thisimagehalfwidth,ymin=-\thisimagehalfheight,xmax=\thisimagehalfwidth,ymax=\thisimagehalfheight] {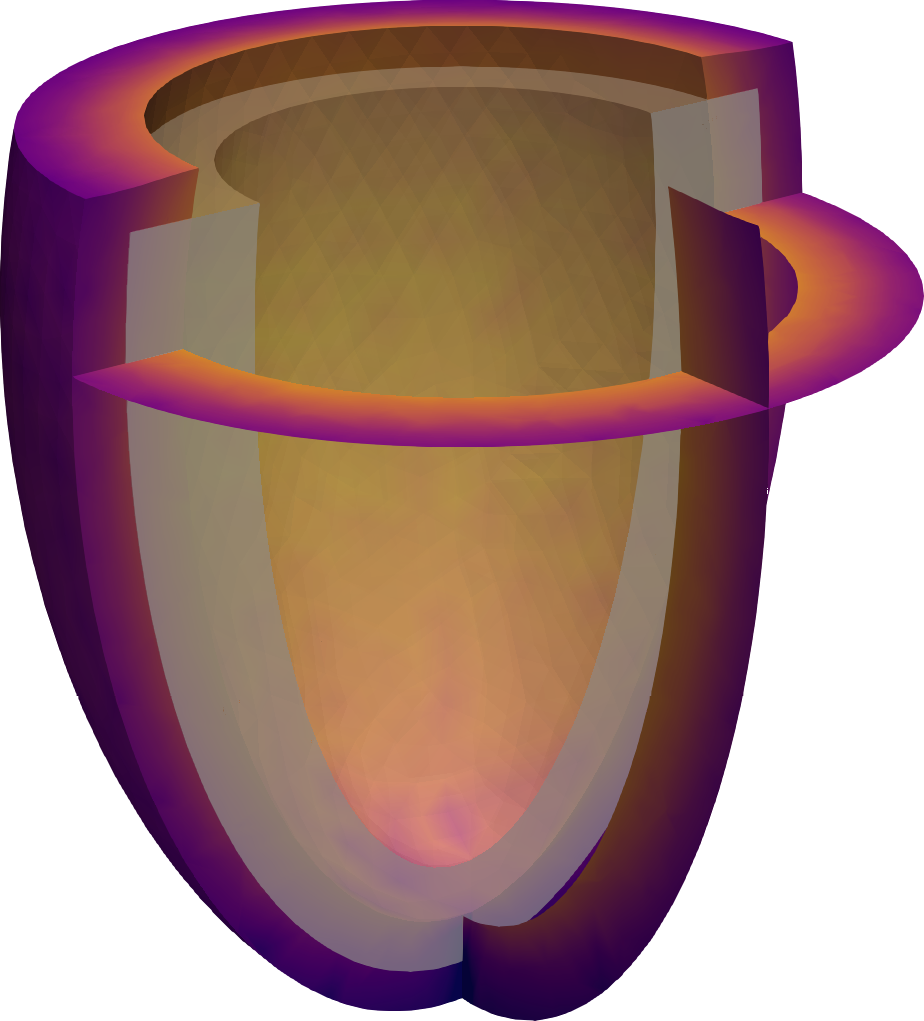};
        \end{axis}

        \import{figures/data/}{case180b-3_gr_last.tex}
        \begin{axis}[
                hide axis,
                width=6.5cm,
                scale only axis,
                enlargelimits=false,
                xmin=-569.5,
                xmax=569.5,
                ymin=-606,
                ymax=606,
                axis equal=true,
                name=plot2,
                at={(plot1.north east)},
                anchor=north west
            ]
            \addplot[thick,blue] graphics[xmin=-\thisimagehalfwidth,ymin=-\thisimagehalfheight,xmax=\thisimagehalfwidth,ymax=\thisimagehalfheight] {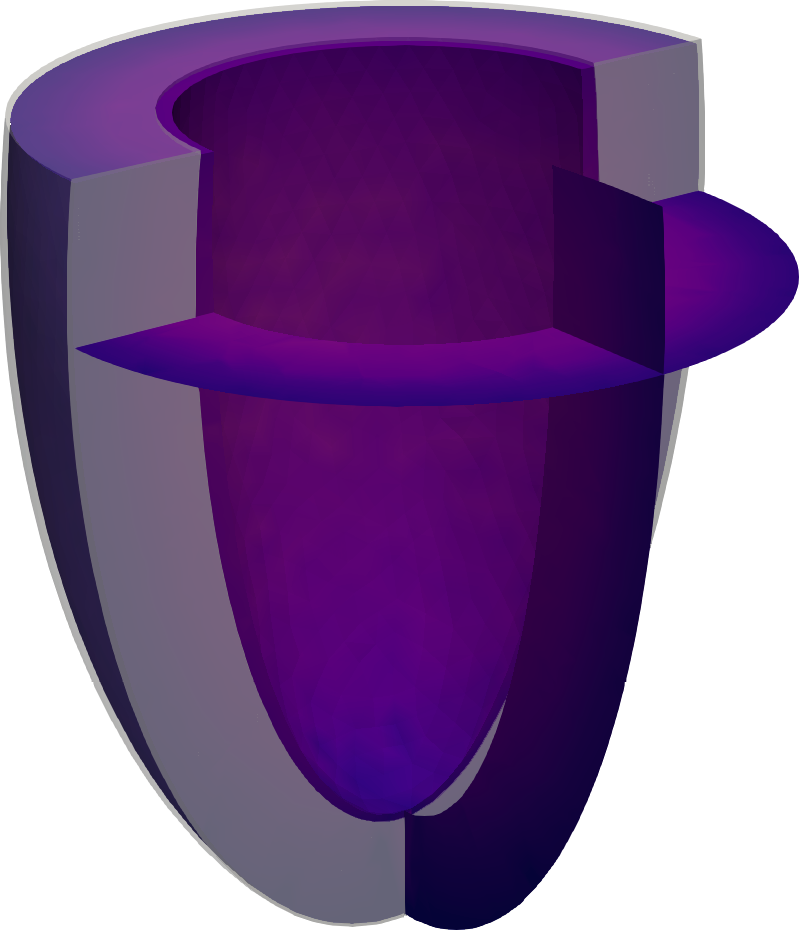};
        \end{axis}
        \begin{axis}[
                hide axis,
                colormap name=thisplotcolormap,
                colorbar,
                colorbar style={%
                    yticklabel={\pgfmathprintnumber\tick\,\%},
                    ylabel=Local mass increase,
                },
                point meta max=\volumechangemax,
                point meta min=\volumechangemin,
                name=plotcolorbar,
                at={(plot2.north east)},
                anchor=north west
            ]
        \end{axis}

        \node[color=dia3x3, draw, above = 0cm of plot1.south, anchor=center] {A};
        \node[color=dia3x3, draw, above = 0cm of plot2.south, anchor=center] {C};

    \end{tikzpicture}
\end{center}
    \caption{Stage 2 hypertension with an early intervention after $150\,\si{days}$.
        The ventricle was loaded with a systolic pressure of $180\,\si{mmHg}$.
        Configuration [A] after $150\,\si{days}$ of G\&R. The shown configuration is not in a homeostatic state. Configuration [C]
        after returning pressure to baseline $p=120\,\si{mmHg}$ and reaching a homeostatic state. Irreversible changes occurred, resulting in an increased wall
        thickness.}
    \label{fig:case180_early_reversal_3d}
\end{figure*}

As shown in Section~\ref{sec:results:stability}, stage 2 hypertension resulted in unstable growth. Here, we
simulated the reversal of G\&R that occurs after an early intervention, more precisely a decrease
of the systolic pressure to the baseline value after $150\,\si{days}$ of continuous G\&R.
We focus on the blue curve in the time interval $s > 150\,\si{days}$. The elastic response of the
pressure normalization results in a sudden decrease of the cardiomyocytes stress to $20.5 \pm 0.5\,\si{kPa}$, a step
increase to a wall thickness of $9.6\,\si{mm} $ and a slight decrease of the endocardial diameter.
In the subsequent G\&R phase, the cardiomyocytes stress exponentially increased towards the ho\-meo\-sta\-tic stress. 
The mid-cavity wall thickness stabilized at $11.8\,\si{mm}$ ($+10\,\si{\%}$ compared to the reference configuration).
Compared to the baseline collagen mass fraction ($10\,\%$), a slightly increased collagen mass fraction is observed $11 \pm 1\,\%$ in the final state.
The endocardial diameter of the final (stable) configuration is $35\,\si{mm}$ ($-8\,\si{\%}$).
The final configuration [C] is depicted in Figure~\ref{fig:case180_early_reversal_3d}, with the reference configuration
in light gray.

\section{Discussion}
\label{sec:discussion}

In this work, we developed a computational model of organ-scale cardiac G\&R that is based on me\-chano\-bio\-logical principles and motivated by myocardial microstructure. We modeled G\&R as a consequence of continuous deposition and degradation of sarcomeres and collagen using the homogenized constrained mixture theory and applied it on an idealized 3D left ventricular geometry. We then analyzed me\-chano\-bio\-logical stability of our G\&R model depending on key me\-chano\-bio\-loical parameters and external loading. Additionally, we demonstrated that our model captures the partial reversal of G\&R if systolic pressure is returned to its baseline value.

\subsection{Prestress}

To start our simulations from a state that is in mechanical and homeostatic equilibrium, we introduced a prestressing algorithm which puts the myocardium in a mechanobiologically equilibrated state for a given ventricular pressure.
We demonstrated that our prestressing algorithm yields a physiologically meaningful distribution of elastin prestretch.

We assumed in this exploratory study that on the macroscopic scale, the preferred mechanical environment of the constituents is defined by mechanical stress. We critically note, however, that the question which quantity is the target of homeostasis remains unclear and part of ongoing research \citep{Eichinger2021b}. Another heuristic assumption in our study was that systolic stress defines the target state while it also remains poorly understood how the homeostatic state should be defined and understood best in highly dynamic systems such as the heart.

\subsection{Mechanobiological stability}
\label{sec:disc:stability}

While cardiac G\&R can be a physiological response to exercise or pregnancy, pathological G\&R plays a crucial role in a variety of cardiac diseases. Adaptive short-term G\&R can stabilize cardiac performance to yield a new compensated equilibrium. Yet, in many patients, adverse long-term G\&R is mechanobiologically unstable and progresses to heart failure. Identifying the hallmarks of heart failure to select the best therapy is an urgent clinical challenge \citep{Kim2018a}. Unfortunately, heart failure lacks robust clinical predictors because the links between biomechanical stimuli and adverse G\&R are still unclear \citep{Opie2006a}. There is a pressing need to identify the stimuli of adverse G\&R and predict the propensity of patients with myocardial infarction to develop heart failure. Computational models of full-heart cardiac G\&R, informed by cardiac magnetic resonance imaging, show high potential to fill this gap and to uncover links between biomechanics and cellular mechanisms in a controlled virtual environment \citep{Niestrawska2020a}.

Previous models of cardiac G\&R based on the kinematic growth theory \emph{a priori} prescribed the direction and extent of growth. \cite{Witzenburg2017a} tested six different cardiac G\&R models based on the kinematic growth theory and found that none was able to return to homeostasis after mixed pressure and volume overload conditions. In our model, the direction and extent of G\&R emerge naturally from intra- and extracellular turnover processes in myocardial tissue constituents. Interpreting the increase in left ventricular pressure as a perturbation of the myocardium’s initial healthy homeostatic state, we demonstrated that our model can predict both mechanobiologically stable and unstable cardiac configurations. Importantly, these configurations can either arise from changes in key parameters (like the product of the mass production parameter $k^i$ and the turnover time $T^i$) or even for an identical model parameter set due to different levels of hypertension. The latter observation is particularly interesting because it extends studies of mechanobiological stability beyond the small perturbation regime studied in \cite{Cyron2014a}. If the pressure perturbation in our system is large enough, the system is pushed out of the stable regime and subsequent G\&R does not converge towards the original or also a new stable mechanobiological equilibrium state but rather continues in an unbounded manner. We can thus identify parameter regions for a patient's heart where it is more likely to return to a stable equilibrium through G\&R. The observation that fast adaption rates can stabilize G\&R was previously made for arterial tissue as well \citep{Latorre2019a}.

\subsection{Reversal}
\label{sec:disc:reversal}

Living tissue such as the myocardium can increase and decrease in size. However, many existing
computational models have focused their modeling on mass increase. If only the pathological
case is of interest, this would be sufficient. However, there are medications and medical devices that aim at stopping or even reversing cardiac G\&R. In clinical practice, it is, thus, of interest if
and how well a patient responds to a certain therapy, which necessitates that reverse
remodeling is also captured by a model.
We demonstrated that our model
can inherently describe also the reversal of G\&R.

We have analyzed the reversal of G\&R in two scenarios in detail. The reversal of a stabilized stage 1 hypertrophy resulted almost in the original configuration. This means that during the preceding
hypertension period, mainly reversible G\&R occurred. In the case of unstable stage 2
hypertension case, we simulated an early intervention after $150\,\si{days}$ following the increased
pressure. Here, the final reversed configuration is distinct from the reversal of stage 1 hypertension: An increased wall thickness remains with a decreased cavity size and a significant amount of fibrosis. The G\&R that occurred in the predecessing unstable hypertension period partially resulted in irreversible G\&R, despite the pressure returning to its baseline value. This is also observed in real hearts: Adaptive growth due to high blood demand (e.g., during exercise or pregnancy) is also termed \emph{reversible growth}, whereas maladaptive growth is termed \emph{irreversible growth}.

\subsection{Phenotype}
\label{sec:phenotype}

In our simulations, hypertension in the left ventricle resulted in dilation of the ventricle with a
reduction in wall thickness -- typically termed eccentric hypertrophy. Left ventricular G\&R
following a pressure overload is often experimentally analyzed with aortic banding experiments \citep{Imamura1990a,Roussel2008a}.
These experiments typically show an increase in wall thickness a few days to weeks after banding, hence, mainly concentric
hypertrophy. In clinical practice, this pattern is less obvious. Typically, a mix of both patterns, concentric
and eccentric hypertrophy can be observed. \cite{Cuspidi2012a} found evidence that eccentric hypertrophy was more
prevalent than concentric hypertrophy in their examined studies. It is important to note that
co-pathologies often influence the observations in clinical practice.

Currently, our model only considers the end-systolic configuration of the heart cycle.
Growth stimuli are, however, active in the whole cardiac cycle. Our model currently cannot detect the difference
between growth stimuli of pressure and volume overload and, therefore, cannot yet represent both concentric
and eccentric hypertrophy.

Constrained mixture type
models consider G\&R of different constituents individually. Our model predicts a significant increase in
collagen mass fraction which is also observed in models of rat left ventricles after 4 weeks of aortic
banding \citep{Doering1988a}. \revc{Our model also predicts a gradient in left ventricle fibrosis from endo- to epicardium
as observed in patients with systemic hypertension \citep{Cowling2019a}.} In the severe hypertensive case, our model predicts only a partial
reversal of fibrosis which is in line with studies on patients after treatment of advanced hypertension \citep{Frangogiannis2020a}.

\subsection{\reva{Experimental validation}}
\label{sec:discussion:validation}

\reva{
    Experimental data for validating microstructural features of our model remain lacking \citep{humphrey2021a}, hence,
    we rely on the qualitative comparisons in Section~\ref{sec:phenotype}. For a more in-depth
    quantitative validation, experimental data on the microstructural and the organ scale are necessary.
}

\reva{
    \cite{cyron2016a} have shown that homogenized constrained mixture models can capture results
    from experiments with tissue equivalents. The results of these experiments, however,
    strongly depend on the composition of the tissue equivalents making it difficult to deduce
    material parameters from these experiments for our organ-scale model. \cite{Fischer2019a} demonstrated how living myocardium
    excised from explanted failing hearts can be analyzed in a biomimetic environment. Analyzing long-term
    G\&R with such an experimental setup might one day allow for calibrating our  G\&R model with species-specific
    parameters of the heart.
}

\reva{
    To validate the model on the organ scale, longitudinal imaging data of cardiac G\&R is necessary.
    \cite{Stoeck2021a} analyzed the response of the myocardium to infarction on a porcine model.
    Their results could be used to verify a model of post-myocardial infarction G\&R. We cannot
    validate our model against their results since modeling of myocardial infarction has
    not yet been done, but it is one of our following projects.
}

\subsection{Computational performance}
The computational cost of homogenized constrained mixture models is comparable to the one of kinematic growth models and much lower than typically for classical constrained mixture models based on multi-network theory \citep{Humphrey2002b}. \cite{Latorre2018a} recently proposed an interesting  approach to speed up computations based on classical con\-strained mixture models that directly yields the long-term outcome of G\&R. However, with this kind of model, reversal of an intermediate configuration (early intervention) is not possible making homogenized constrained mixture models the best available trade-off between computational cost and physiological realism for our study.

The computational costs of the prestressing algorithm are comparable to the one of the successive G\&R simulation. However, prestressing has to be done only once per geometry and the resulting configuration can be used to simulate multiple G\&R scenarios.

With our, so far, not optimized implementation, we ran up to 8 different G\&R scenarios on one Intel Xeon E5-2630 v3 ``Haswell'' (16 cores, $2.5\,\si{GHz}$, $64\,\si{GB}$) of our Linux cluster. The prestress algorithm needed around $33\,\si{min}$ ($\varepsilon_\text{pre}=0.01\,\si{mm}$) to obtain a mechanobiological equilibrated reference configuration. The prestress algorithm had on average $3.2$ Newton iterations per step and $219$ linear solver iterations per Newton iteration. The subsequent G\&R took around $1\,\si{h}$ for the first $250\,\si{days}$ of G\&R. Stage 1 hypertension had on average $3.4$ Newton iterations per time step and $273$ linear solver iterations per Newton iteration with a downward tendency while approaching homeostasis. For Stage 2 hypertension, these values were $3.9$ and $389$, respectively, with a tendency to rise during runaway growth.

\subsection{\revc{Limitations and} future perspectives}
\label{sec:future_perspectives}

Our simulations were conducted on a truncated spheroid as an approximation of the left ventricle
of the heart. The geometry of the heart also influences G\&R in the ventricles. In general,
homogenized constrained mixture models are not limited to regular geometries. \cite{Mousavi2019a}
analyzed the evolution of ascending thoracic aortic aneurysms on patient-specific geometries
using the homogenized constrained mixture model. A computational model of cardiac G\&R that
bases on the homogenized constrained mixture model on patient-specific geometries is one of our next
steps towards the goal of clinical applications.

We only differentiated between me\-chano\-bio\-logi\-cally stable and unstable G\&R. However, there are cases where the shape change is too deleterious, and heart failure occurs despite G\&R reaching a stable limit, or in contrast, cases where unstable G\&R occurs on a very long time scale so that the deleterious shape change
would occur long after the natural death of the patient. A multi-timescale simulation with the G\&R scale and the heart cycle scale is necessary to capture the transient effects of heart failure. Here, the simulation of several cardiac contractions could be sped up with reduced-order models \citep{Pfaller2020a}.

We describe myocardial tissue as a constrained mixture of structurally relevant constituents, each having a different prestretch. Material parameters for this class of models for the myocardium are lacking. Standard material models account only for the overall stress response without incorporating different constituents and their prestretch. Ex-vivo experimental data on myocardial tissue \citep{Sommer2015a} can be used to identify a set of parameters of a prestressed constrained mixture. %

Currently, we use the same model for turnover of cardiomyocytes and collagen fibers. Constrained
mixture type models can, however, use individual models for different constituents. Since turnover
in cardiomyocytes is an intracellular process \citep{Willis2009a}, the microstructural processes
differ from the observations of collagen turnover. Individual models for each constituent could be
developed which are based on constituent-specific assumptions and microstructural observations.

\reva{There are currently no reliable models for entropy influx
during G\&R of soft tissue. As a consequence, we cannot discuss the thermodynamic consistency of the model in a strong sense.
Constraints imposed by the Clausius-Duhem inequality (second law of thermodynamics) need an increased theoretical understanding of soft tissue G\&R \citep{humphrey2021a}.}

Currently, we assumed that only the myocardial tissue itself adapts to changed loads. We
assumed that the behavior of the pericardial boundary condition, the cut-off of the atria
and the hemodynamic loads do not change over time. \revc{However, the pericardium is known to enlarge
in size and shape during cardiac dilatation \citep{Freeman1984a} and the atria will also adapt to
changes in the mechanical environment.} To model the adaption of the atria and the
right ventricle, a four-chamber geometry of the heart could be used. Modeling the change
in hemodynamics is however more involved. Models have been developed to describe
the regulation of blood pressure on the long timescale \citep{Beard2013a}. Recently,
\cite{Pourmodheji2022a} coupled right ventricular and pulmonary arterial G\&R in a multi-temporal
computational model.

As discussed in Section~\ref{sec:phenotype}, cardiac G\&R concerns two time-scales, G\&R (days) vs.
a single heartbeat (seconds). In our work thus far, we have only considered the long-term G\&R scale.
However, biomechanical loads on myocardial constituents change significantly over one cardiac
cycle. In future work, we plan to extract G\&R stimuli from the transient load through the cardiac
cycle. This will allow our model to better distinguish between the G\&R patterns of concentric and
eccentric hypertrophy.

\revb{
    A global sensitivity analysis \citep{Brandstaeter2021a} of our model
    could identify parameters with a high influence on mechanobiological stability. This could provide
    answers for the clinically relevant question of whether a patient with specific features (shape of
    the heart, composition of the myocardium, overload conditions, ...) is at risk of unstable G\&R.
}

\section*{Declarations}

\section*{Ethical Approval}
not applicable

\section*{Competing interests}
The authors declare that they have no competing interests.

\section*{Authors' contributions}
\begin{itemize}
    \item AMG: Methodology, Software, Validation, Visualization, Writing –- original draft.
    \item MRP: Funding acquisition, Methodology, Validation, Visualization, Writing –- review \& editing.
    \item FAB: Methodology, Software, Visualization, Writing –- review \& editing.
    \item CJC: Methodology, Writing –- review \& editing.
    \item WAW: Conceptualization, Funding acquisition, Resources, Supervision, Writing -– review \& editing.
\end{itemize}

\section*{Funding}
MRP acknowledges the support by the National Heart, Lung, and Blood Institute
of the National Institutes of Health under Award Number K99HL161313 and the Stanford Maternal
and Child Health Research Institute. The content is solely the responsibility of the authors and does not necessarily represent the official views of the National Institutes of Health.
WAW was supported by BREATHE, a Horizon 2020—ERC-2020-ADG project, grant agreement No. 101021526-BREATHE.

\section*{Availability of data and materials}
All data generated or analyzed during this study are included in this published article.

\bibliographystyle{spbasic}      %
\bibliography{literature}

\begin{thebibliography}{68}
\providecommand{\natexlab}[1]{#1}
\providecommand{\url}[1]{{#1}}
\providecommand{\urlprefix}{URL }
\expandafter\ifx\csname urlstyle\endcsname\relax
  \providecommand{\doi}[1]{DOI~\discretionary{}{}{}#1}\else
  \providecommand{\doi}{DOI~\discretionary{}{}{}\begingroup
  \urlstyle{rm}\Url}\fi
\providecommand{\eprint}[2][]{\url{#2}}

\bibitem[{Aboelkassem et~al.(2019)Aboelkassem, Powers, McCabe, and
  McCulloch}]{Aboelkassem2019a}
Aboelkassem Y, Powers JD, McCabe KJ, McCulloch AD (2019) {Multiscale Models of
  Cardiac Muscle Biophysics and Tissue Remodeling in Hypertrophic
  Cardiomyopathies}. Current Opinion in Biomedical Engineering 11:35--44,
  \doi{10.1016/j.cobme.2019.09.005}

\bibitem[{{BACI}(2021)}]{BACI}
{BACI} (2021) {BACI}: A comprehensive multi-physics simulation framework.
  \url{https://baci.pages.gitlab.lrz.de/website}, accessed: 2021-07-28

\bibitem[{Beard et~al.(2013)Beard, Pettersen, Carlson, Omholt, and
  Bugenhagen}]{Beard2013a}
Beard DA, Pettersen KH, Carlson BE, Omholt SW, Bugenhagen SM (2013) {A
  computational analysis of the long-term regulation of arterial pressure}.
  F1000Research 2:208, \doi{10.12688/f1000research.2-208.v1}

\bibitem[{Bergmann et~al.(2009)Bergmann, Bhardwaj, Bernard, Zdunek,
  Barnabé-Heider, Walsh, Zupicich, Alkass, Buchholz, Druid, Jovinge, and
  Frisén}]{Bergmann2009a}
Bergmann O, Bhardwaj RD, Bernard S, Zdunek S, Barnabé-Heider F, Walsh S,
  Zupicich J, Alkass K, Buchholz BA, Druid H, Jovinge S, Frisén J (2009)
  {Evidence for Cardiomyocyte Renewal in Humans}. Science 324(5923):98--102,
  \doi{10.1126/science.1164680}

\bibitem[{Bergmann et~al.(2015)Bergmann, Zdunek, Felker, Salehpour, Alkass,
  Bernard, Sjostrom, Szewczykowska, Jackowska, dos Remedios, Malm, Andrä,
  Jashari, Nyengaard, Possnert, Jovinge, Druid, and Frisén}]{Bergmann2015a}
Bergmann O, Zdunek S, Felker A, Salehpour M, Alkass K, Bernard S, Sjostrom S,
  Szewczykowska M, Jackowska T, dos Remedios C, Malm T, Andrä M, Jashari R,
  Nyengaard J, Possnert G, Jovinge S, Druid H, Frisén J (2015) {Dynamics of
  Cell Generation and Turnover in the Human Heart}. Cell 161(7):1566--1575,
  \doi{10.1016/j.cell.2015.05.026}

\bibitem[{Braeu et~al.(2016)Braeu, Seitz, Aydin, and Cyron}]{braeu2016a}
Braeu FA, Seitz A, Aydin RC, Cyron CJ (2016) {Homogenized constrained mixture
  models for anisotropic volumetric growth and remodeling}. Biomechanics and
  Modeling in Mechanobiology 16(3):889--906, \doi{10.1007/s10237-016-0859-1}

\bibitem[{Braeu et~al.(2019)Braeu, Aydin, and Cyron}]{braeu2019a}
Braeu FA, Aydin RC, Cyron CJ (2019) {Anisotropic stiffness and tensional
  homeostasis induce a natural anisotropy of volumetric growth and remodeling
  in soft biological tissues}. Biomechanics and Modeling in Mechanobiology
  18(2):327--345, \doi{10.1007/s10237-018-1084-x}

\bibitem[{Brandstaeter et~al.(2021)Brandstaeter, Fuchs, Biehler, Aydin, Wall,
  and Cyron}]{Brandstaeter2021a}
Brandstaeter S, Fuchs SL, Biehler J, Aydin RC, Wall WA, Cyron CJ (2021) {Global
  Sensitivity Analysis of a Homogenized Constrained Mixture Model of Arterial
  Growth and Remodeling}. Journal of Elasticity pp 1--31,
  \doi{10.1007/s10659-021-09833-9}

\bibitem[{Brower et~al.(2006)Brower, Gardner, Forman, Murray, Voloshenyuk,
  Levick, and Janicki}]{Brower2006a}
Brower GL, Gardner JD, Forman MF, Murray DB, Voloshenyuk T, Levick SP, Janicki
  JS (2006) {The relationship between myocardial extracellular matrix
  remodeling and ventricular function}. European Journal of Cardio-Thoracic
  Surgery 30(4):604--610, \doi{10.1016/j.ejcts.2006.07.006}

\bibitem[{Brown et~al.(1998)Brown, Prajapati, McGrouther, Yannas, and
  Eastwood}]{Brown1998a}
Brown RA, Prajapati R, McGrouther DA, Yannas IV, Eastwood M (1998) {Tensional
  homeostasis in dermal fibroblasts: Mechanical responses to mechanical loading
  in three‐dimensional substrates}. Journal of Cellular Physiology
  175(3):323--332,
  \doi{10.1002/(sici)1097-4652(199806)175:3<323::aid-jcp10>3.0.co;2-6}

\bibitem[{Chobanian et~al.(2003)Chobanian, Bakris, and Black}]{Chobanian2003a}
Chobanian A, Bakris G, Black H (2003) The seventh report of the joint national
  committee on prevention, detection, evaluation, and treatment of high blood
  pressure. the {JNC} 7 report. {ACC} Current Journal Review 12(4):31--32,
  \doi{10.1016/s1062-1458(03)00270-8}

\bibitem[{Cocciolone et~al.(2018)Cocciolone, Hawes, Staiculescu, Johnson,
  Murshed, and Wagenseil}]{Cocciolone2018a}
Cocciolone AJ, Hawes JZ, Staiculescu MC, Johnson EO, Murshed M, Wagenseil JE
  (2018) {Elastin, arterial mechanics, and cardiovascular disease}. American
  Journal of Physiology-Heart and Circulatory Physiology 315(2):H189--H205,
  \doi{10.1152/ajpheart.00087.2018}

\bibitem[{Cohn et~al.(2000)Cohn, Ferrari, and Sharpe}]{Cohn2000a}
Cohn JN, Ferrari R, Sharpe N (2000) Cardiac remodeling—concepts and clinical
  implications: a consensus paper from an international forum on cardiac
  remodeling. Journal of the American College of Cardiology 35(3):569--582,
  \doi{10.1016/S0735-1097(99)00630-0}

\bibitem[{Cowling et~al.(2019)Cowling, Kupsky, Kahn, Daniels, and
  Greenberg}]{Cowling2019a}
Cowling RT, Kupsky D, Kahn AM, Daniels LB, Greenberg BH (2019) {Mechanisms of
  cardiac collagen deposition in experimental models and human disease}.
  Translational Research 209:138--155, \doi{10.1016/j.trsl.2019.03.004}

\bibitem[{Cuspidi et~al.(2012)Cuspidi, , Sala, Negri, Mancia, and
  Morganti}]{Cuspidi2012a}
Cuspidi C, , Sala C, Negri F, Mancia G, Morganti A (2012) {Prevalence of
  left-ventricular hypertrophy in hypertension: an updated review of
  echocardiographic studies}. Journal of Human Hypertension 26(6):343--349,
  \doi{10.1038/jhh.2011.104}

\bibitem[{Cyron and Hum\-phrey(2014)}]{Cyron2014a}
Cyron CJ, Hum\-phrey JD (2014) {Vascular homeostasis and the concept of
  mechanobiological stability}. International Journal of Engineering Science
  85:203--223, \doi{10.1016/j.ijengsci.2014.08.003}

\bibitem[{Cyron and Humphrey(2017)}]{Cyron2017a}
Cyron CJ, Humphrey JD (2017) {Growth and remodeling of load-bearing biological
  soft tissues}. Meccanica 52(3):645--664, \doi{10.1007/s11012-016-0472-5}

\bibitem[{Cyron et~al.(2014)Cyron, Wilson, and Humphrey}]{Cyron2014b}
Cyron CJ, Wilson JS, Humphrey JD (2014) {Mechanobiological stability: a new
  paradigm to understand the enlargement of aneurysms?} Journal of The Royal
  Society Interface 11(100):20140680, \doi{10.1098/rsif.2014.0680}

\bibitem[{Cyron et~al.(2016)Cyron, Aydin, and Humphrey}]{cyron2016a}
Cyron CJ, Aydin RC, Humphrey JD (2016) {A homogenized constrained mixture (and
  mechanical analog) model for growth and remodeling of soft tissue}.
  Biomechanics and Modeling in Mechanobiology 15(6):1389--1403,
  \doi{10.1007/s10237-016-0770-9}

\bibitem[{Doering et~al.(1988)Doering, Jalil, Janicki, Pick, Aghili, Abrahams,
  and Weber}]{Doering1988a}
Doering CW, Jalil JE, Janicki JS, Pick R, Aghili S, Abrahams C, Weber KT (1988)
  {Collagen network remodelling and diastolic stiffness of the rat left
  ventricle with pressure overload hypertrophy}. Cardiovascular Research
  22(10):686--695, \doi{10.1093/cvr/22.10.686}

\bibitem[{Eichinger et~al.(2021{\natexlab{a}})Eichinger, Haeusel, Paukner,
  Aydin, Humphrey, and Cyron}]{Eichinger2021a}
Eichinger JF, Haeusel LJ, Paukner D, Aydin RC, Humphrey JD, Cyron CJ
  (2021{\natexlab{a}}) {Mechanical homeostasis in tissue equivalents: a
  review}. Biomechanics and Modeling in Mechanobiology 20(3):833--850,
  \doi{10.1007/s10237-021-01433-9}

\bibitem[{Eichinger et~al.(2021{\natexlab{b}})Eichinger, Paukner, Aydin, Wall,
  Humphrey, and Cyron}]{Eichinger2021b}
Eichinger JF, Paukner D, Aydin RC, Wall WA, Humphrey JD, Cyron CJ
  (2021{\natexlab{b}}) {What do cells regulate in soft tissues on short time
  scales?} Acta Biomaterialia 134:348--356, \doi{10.1016/j.actbio.2021.07.054},
  \eprint{2104.05580}

\bibitem[{Estrada et~al.(2020)Estrada, Yoshida, Saucerman, and
  Holmes}]{Estrada2020a}
Estrada AC, Yoshida K, Saucerman JJ, Holmes JW (2020) {A multiscale model of
  cardiac concentric hypertrophy incorporating both mechanical and hormonal
  drivers of growth}. Biomechanics and Modeling in Mechanobiology pp 1--15,
  \doi{10.1007/s10237-020-01385-6}

\bibitem[{Ezra et~al.(2010)Ezra, Ellis, Beaconsfield, Collin, and
  Bailly}]{Ezra2010a}
Ezra DG, Ellis JS, Beaconsfield M, Collin R, Bailly M (2010) {Changes in
  Fibroblast Mechanostat Set Point and Mechanosensitivity: An Adaptive Response
  to Mechanical Stress in Floppy Eyelid Syndrome}. Investigative Ophthalmology
  \& Visual Science 51(8):3853--3863, \doi{10.1167/iovs.09-4724}

\bibitem[{Fan et~al.(2021)Fan, Coll-Font, Boomen, Kim, Chen, Eder, Roche, and
  Nguyen}]{Fan2021a}
Fan Y, Coll-Font J, Boomen Mvd, Kim JH, Chen S, Eder RA, Roche ET, Nguyen CT
  (2021) {Characterization of Exercise-Induced Myocardium Growth Using Finite
  Element Modeling and Bayesian Optimization}. Frontiers in Physiology
  12:694940, \doi{10.3389/fphys.2021.694940}

\bibitem[{Fischer et~al.(2019)Fischer, Milting, Fein, Reiser, Lu, Seidel,
  Schinner, Schwarzmayr, Schramm, Tomasi, Husse, Cao-Ehlker, Pohl, and
  Dendorfer}]{Fischer2019a}
Fischer C, Milting H, Fein E, Reiser E, Lu K, Seidel T, Schinner C, Schwarzmayr
  T, Schramm R, Tomasi R, Husse B, Cao-Ehlker X, Pohl U, Dendorfer A (2019)
  {Long-term functional and structural preservation of precision-cut human
  myocardium under continuous electromechanical stimulation in vitro}. Nature
  Communications 10(1):117, \doi{10.1038/s41467-018-08003-1}

\bibitem[{Frangogiannis(2020)}]{Frangogiannis2020a}
Frangogiannis NG (2020) {Cardiac fibrosis}. Cardiovascular Research
  117(6):1450--1488, \doi{10.1093/cvr/cvaa324}

\bibitem[{Freeman and LeWinter(1984)}]{Freeman1984a}
Freeman GL, LeWinter MM (1984) {Pericardial adaptations during chronic cardiac
  dilation in dogs.} Circulation Research 54(3):294--300,
  \doi{10.1161/01.res.54.3.294}

\bibitem[{Geuzaine and Remacle(2009)}]{Geuzaine2009a}
Geuzaine C, Remacle J (2009) {Gmsh: A 3‐D finite element mesh generator with
  built‐in pre‐ and post‐processing facilities}. International Journal
  for Numerical Methods in Engineering 79(11):1309--1331,
  \doi{10.1002/nme.2579}

\bibitem[{Grossman et~al.(1975)Grossman, Jones, and McLaurin}]{Grossman1975a}
Grossman W, Jones D, McLaurin LP (1975) {Wall stress and patterns of
  hypertrophy in the human left ventricle.} Journal of Clinical Investigation
  56(1):56--64, \doi{10.1172/jci108079}

\bibitem[{Göktepe et~al.(2010)Göktepe, Abilez, Parker, and
  Kuhl}]{Goektepe2010a}
Göktepe S, Abilez OJ, Parker KK, Kuhl E (2010) {A multiscale model for
  eccentric and concentric cardiac growth through sarcomerogenesis}. Journal of
  Theoretical Biology 265(3):433--442, \doi{10.1016/j.jtbi.2010.04.023}

\bibitem[{Holzapfel(2000)}]{holzapfel2000a}
Holzapfel GA (2000) {Nonlinear Solid Mechanics: A Continuum Approach for
  Enineering}. Wiley

\bibitem[{Holzapfel and Ogden(2009)}]{Holzapfel2009a}
Holzapfel GA, Ogden RW (2009) {Constitutive modelling of passive myocardium: a
  structurally based framework for material characterization}. Philosophical
  Transactions of the Royal Society A: Mathematical, Physical and Engineering
  Sciences 367(1902):3445--3475, \doi{10.1098/rsta.2009.0091}

\bibitem[{Humphrey(2002)}]{Humphrey2002b}
Humphrey J (2002) Cardiovascular Solid Mechanics: Cells, Tissues, and Organs,
  vol~55. \doi{10.1115/1.1497492}

\bibitem[{Humphrey(2021)}]{humphrey2021a}
Humphrey JD (2021) {Constrained Mixture Models of Soft Tissue Growth and
  Remodeling – Twenty Years After}. Journal of Elasticity pp 1--27,
  \doi{10.1007/s10659-020-09809-1}

\bibitem[{Humphrey and Rajagopal(2002)}]{humphrey2002a}
Humphrey JD, Rajagopal KR (2002) {A Constrained Mixture Model for Growth and
  Remodeling of Soft Tissues}. Mathematical Models and Methods in Applied
  Sciences 12(03):407--430, \doi{10.1142/s0218202502001714}

\bibitem[{Imamura et~al.(1990)Imamura, Schluchter, and
  Fouad-Tarazi}]{Imamura1990a}
Imamura M, Schluchter M, Fouad-Tarazi FM (1990) {Remodelling of left ventricle
  after banding of ascending aorta in the rat}. Cardiovascular Research
  24(8):641--646, \doi{10.1093/cvr/24.8.641}

\bibitem[{Kehat and Molkentin(2010)}]{Kehat2010a}
Kehat I, Molkentin JD (2010) {Molecular Pathways Underlying Cardiac Remodeling
  During Pathophysiological Stimulation}. Circulation 122(25):2727--2735,
  \doi{10.1161/circulationaha.110.942268}

\bibitem[{Kim et~al.(2018)Kim, Uriel, and Burkhoff}]{Kim2018a}
Kim GH, Uriel N, Burkhoff D (2018) {Reverse remodelling and myocardial recovery
  in heart failure}. Nature Reviews Cardiology 15(2):83--96,
  \doi{10.1038/nrcardio.2017.139}

\bibitem[{Kroon et~al.(2009)Kroon, Delhaas, Arts, and Bovendeerd}]{Kroon2009a}
Kroon W, Delhaas T, Arts T, Bovendeerd P (2009) {Computational modeling of
  volumetric soft tissue growth: application to the cardiac left ventricle}.
  Biomechanics and Modeling in Mechanobiology 8(4):301--309,
  \doi{10.1007/s10237-008-0136-z}

\bibitem[{Latorre and Humphrey(2018)}]{Latorre2018a}
Latorre M, Humphrey JD (2018) {A mechanobiologically equilibrated constrained
  mixture model for growth and remodeling of soft tissues}. ZAMM - Journal of
  Applied Mathematics and Mechanics / Zeitschrift für Angewandte Mathematik
  und Mechanik 98(12):2048--2071, \doi{10.1002/zamm.201700302}

\bibitem[{Latorre and Humphrey(2019)}]{Latorre2019a}
Latorre M, Humphrey JD (2019) {Mechanobiological stability of biological soft
  tissues}. Journal of the Mechanics and Physics of Solids 125:298--325,
  \doi{10.1016/j.jmps.2018.12.013}

\bibitem[{Lee et~al.(2016)Lee, Kassab, and Guccione}]{Lee2016b}
Lee L, Kassab G, Guccione J (2016) {Mathematical modeling of cardiac growth and
  remodeling}. Wiley Interdisciplinary Reviews: Systems Biology and Medicine
  8(3):211--226, \doi{10.1002/wsbm.1330}

\bibitem[{Lee et~al.(2015)Lee, Genet, Acevedo-Bolton, Ordovas, Guccione, and
  Kuhl}]{Lee2015a}
Lee LC, Genet M, Acevedo-Bolton G, Ordovas K, Guccione JM, Kuhl E (2015) {A
  computational model that predicts reverse growth in response to mechanical
  unloading}. Biomechanics and Modeling in Mechanobiology 14(2):217--229,
  \doi{10.1007/s10237-014-0598-0}

\bibitem[{Linzbach(1960)}]{Linzbach1960a}
Linzbach A (1960) {Heart failure from the point of view of quantitative
  anatomy}. The American Journal of Cardiology 5(3):370--382,
  \doi{10.1016/0002-9149(60)90084-9}

\bibitem[{Mousavi and Avril(2017)}]{mousavi2017a}
Mousavi SJ, Avril S (2017) {Patient-specific stress analyses in the ascending
  thoracic aorta using a finite-element implementation of the constrained
  mixture theory.} Biomechanics and modeling in mechanobiology
  16(5):1765--1777, \doi{10.1007/s10237-017-0918-2}

\bibitem[{Mousavi et~al.(2019)Mousavi, Farzaneh, and Avril}]{Mousavi2019a}
Mousavi SJ, Farzaneh S, Avril S (2019) {Patient-specific predictions of
  aneurysm growth and remodeling in the ascending thoracic aorta using the
  homogenized constrained mixture model}. Biomechanics and Modeling in
  Mechanobiology 18(6):1895--1913, \doi{10.1007/s10237-019-01184-8},
  \eprint{1912.07884}

\bibitem[{Nagler et~al.(2017)Nagler, Bertoglio, Stoeck, Kozerke, and
  Wall}]{Nagler2017a}
Nagler A, Bertoglio C, Stoeck CT, Kozerke S, Wall WA (2017) {Maximum likelihood
  estimation of cardiac fiber bundle orientation from arbitrarily spaced
  diffusion weighted images}. Medical Image Analysis 39:56--77,
  \doi{10.1016/j.media.2017.03.005}

\bibitem[{Niestrawska et~al.(2020)Niestrawska, Augustin, and
  Plank}]{Niestrawska2020a}
Niestrawska JA, Augustin CM, Plank G (2020) {Computational Modeling of Cardiac
  Growth and Remodeling in Pressure Overloaded Hearts — Linking
  Microstructure to Organ Phenotype}. Acta Biomaterialia
  \doi{10.1016/j.actbio.2020.02.010}

\bibitem[{Opie et~al.(2006)Opie, Commerford, Gersh, and Pfeffer}]{Opie2006a}
Opie LH, Commerford PJ, Gersh BJ, Pfeffer MA (2006) {Controversies in
  ventricular remodelling}. The Lancet 367(9507):356--367,
  \doi{10.1016/s0140-6736(06)68074-4}

\bibitem[{Pfaller et~al.(2019)Pfaller, Hörmann, Weigl, Nagler, Chabiniok,
  Bertoglio, and Wall}]{pfaller2019a}
Pfaller MR, Hörmann JM, Weigl M, Nagler A, Chabiniok R, Bertoglio C, Wall WA
  (2019) {The importance of the pericardium for cardiac biomechanics: from
  physiology to computational modeling}. Biomechanics and Modeling in
  Mechanobiology 18(2):503--529, \doi{10.1007/s10237-018-1098-4},
  \eprint{1810.05451}

\bibitem[{Pfaller et~al.(2020)Pfaller, Varona, Lang, Bertoglio, and
  Wall}]{Pfaller2020a}
Pfaller MR, Varona MC, Lang J, Bertoglio C, Wall WA (2020) {Using parametric
  model order reduction for inverse analysis of large nonlinear cardiac
  simulations}. International Journal for Numerical Methods in Biomedical
  Engineering 36(4), \doi{10.1002/cnm.3320}, \eprint{1810.12033}

\bibitem[{Pourmodheji et~al.(2022)Pourmodheji, Jiang, Tossas-Betancourt,
  Dorfman, Figueroa, Baek, and Lee}]{Pourmodheji2022a}
Pourmodheji R, Jiang Z, Tossas-Betancourt C, Dorfman AL, Figueroa CA, Baek S,
  Lee LC (2022) {Computational modelling of multi-temporal
  ventricular–vascular interactions during the progression of pulmonary
  arterial hypertension}. Journal of the Royal Society Interface
  19(196):20220534, \doi{10.1098/rsif.2022.0534}

\bibitem[{Rodriguez et~al.(1994)Rodriguez, Hoger, and
  McCulloch}]{Rodriguez1994a}
Rodriguez EK, Hoger A, McCulloch AD (1994) {Stress-dependent finite growth in
  soft elastic tissues}. Journal of Biomechanics 27(4):455--467,
  \doi{10.1016/0021-9290(94)90021-3}

\bibitem[{Roussel et~al.(2008)Roussel, Gaudreau, Plante, Drolet, Breault,
  Couet, and Arsenault}]{Roussel2008a}
Roussel {\'E}, Gaudreau M, Plante {\'E}, Drolet MC, Breault C, Couet J,
  Arsenault M (2008) {Early responses of the left ventricle to pressure
  overload in Wistar rats}. Life Sciences 82(5-6):265--272,
  \doi{10.1016/j.lfs.2007.11.008}

\bibitem[{Schirone et~al.(2017)Schirone, Forte, Palmerio, Yee, Nocella,
  Angelini, Pagano, Schiavon, Bordin, Carrizzo, Vecchione, Valenti, Chimenti,
  Falco, Sciarretta, and Frati}]{Schirone2017a}
Schirone L, Forte M, Palmerio S, Yee D, Nocella C, Angelini F, Pagano F,
  Schiavon S, Bordin A, Carrizzo A, Vecchione C, Valenti V, Chimenti I, Falco
  ED, Sciarretta S, Frati G (2017) {A Review of the Molecular Mechanisms
  Underlying the Development and Progression of Cardiac Remodeling}. Oxidative
  Medicine and Cellular Longevity 2017:3920195, \doi{10.1155/2017/3920195}

\bibitem[{Sharifi et~al.(2021)Sharifi, Mann, Rockward, Mehri, Mojumder, Lee,
  Campbell, and Wenk}]{Sharifi2021a}
Sharifi H, Mann CK, Rockward AL, Mehri M, Mojumder J, Lee LC, Campbell KS, Wenk
  JF (2021) {Multiscale simulations of left ventricular growth and remodeling}.
  Biophysical Reviews 13(5):729--746, \doi{10.1007/s12551-021-00826-5}

\bibitem[{Sommer et~al.(2015)Sommer, Schriefl, Andrä, Sacherer, Viertler,
  Wolinski, and Holzapfel}]{Sommer2015a}
Sommer G, Schriefl AJ, Andrä M, Sacherer M, Viertler C, Wolinski H, Holzapfel
  GA (2015) {Biomechanical properties and microstructure of human ventricular
  myocardium}. Acta Biomaterialia 24:172--192,
  \doi{10.1016/j.actbio.2015.06.031}

\bibitem[{Spi\-nale(2007)}]{Spinale2007a}
Spi\-nale FG (2007) {Myocardial Matrix Remodeling and the Matrix
  Metalloproteinases: Influence on Cardiac Form and Function}. Physiological
  Reviews 87(4):1285--1342, \doi{10.1152/physrev.00012.2007}

\bibitem[{Stoeck et~al.(2021)Stoeck, Deuster, Fuetterer, Polacin, Waschkies,
  Gorkum, Kron, Fleischmann, Cesarovic, Weisskopf, and Kozerke}]{Stoeck2021a}
Stoeck CT, Deuster Cv, Fuetterer M, Polacin M, Waschkies CF, Gorkum RJHv, Kron
  M, Fleischmann T, Cesarovic N, Weisskopf M, Kozerke S (2021) {Cardiovascular
  magnetic resonance imaging of functional and microstructural changes of the
  heart in a longitudinal pig model of acute to chronic myocardial infarction}.
  Journal of Cardiovascular Magnetic Resonance 23(1):103,
  \doi{10.1186/s12968-021-00794-5}

\bibitem[{{The {T}rilinos {P}roject {T}eam}(2021)}]{trilinos-website}
{The {T}rilinos {P}roject {T}eam} (2021) {The {T}rilinos {P}roject {W}ebsite}.
  \url{{https://trilinos.github.io}}, accessed: 2021-12-09

\bibitem[{Weber(1989)}]{Weber1989a}
Weber KT (1989) {Cardiac interstitium in health and disease: The fibrillar
  collagen network}. Journal of the American College of Cardiology
  13(7):1637--1652, \doi{10.1016/0735-1097(89)90360-4}

\bibitem[{Weisbecker et~al.(2014)Weisbecker, Pierce, and
  Holzapfel}]{weisbecker2014a}
Weisbecker H, Pierce DM, Holzapfel GA (2014) {A generalized prestressing
  algorithm for finite element simulations of preloaded geometries with
  application to the aorta: A generalized prestressing algorithm for finite
  element simulations}. International Journal for Numerical Methods in
  Biomedical Engineering 30(9):857--872, \doi{10.1002/cnm.2632}

\bibitem[{Willis et~al.(2009)Willis, Schisler, Portbury, and
  Patterson}]{Willis2009a}
Willis MS, Schisler JC, Portbury AL, Patterson C (2009) {Build it up–Tear it
  down: protein quality control in the cardiac sarcomere}. Cardiovascular
  Research 81(3):439--448, \doi{10.1093/cvr/cvn289}

\bibitem[{Wilson et~al.(2012)Wilson, Baek, and Humphrey}]{Wilson2012a}
Wilson JS, Baek S, Humphrey JD (2012) {Importance of initial aortic properties
  on the evolving regional anisotropy, stiffness and wall thickness of human
  abdominal aortic aneurysms}. Journal of The Royal Society Interface
  9(74):2047--2058, \doi{10.1098/rsif.2012.0097}

\bibitem[{Witzenburg and Holmes(2017)}]{Witzenburg2017a}
Witzenburg CM, Holmes JW (2017) {A Comparison of Phenomenologic Growth Laws for
  Myocardial Hypertrophy}. Journal of Elasticity 129(1-2):257--281,
  \doi{10.1007/s10659-017-9631-8}

\bibitem[{Yang et~al.(2016)Yang, Schmidt, Wang, Yang, Shao, Borg, Markwald,
  Runyan, and Gao}]{Yang2016a}
Yang H, Schmidt LP, Wang Z, Yang X, Shao Y, Borg TK, Markwald R, Runyan R, Gao
  BZ (2016) {Dynamic Myofibrillar Remodeling in Live Cardiomyocytes under
  Static Stretch}. Scientific Reports 6(1):20674, \doi{10.1038/srep20674}

\bibitem[{Yoshida and Holmes(2020)}]{Yoshida2020a}
Yoshida K, Holmes JW (2020) {Computational models of cardiac hypertrophy}.
  Progress in Biophysics and Molecular Biology
  \doi{10.1016/j.pbiomolbio.2020.07.001}

\end{thebibliography}

\end{document}

\pagenumbering{gobble}

\Large
\begin{center}
    Supplementary material for: A homogenized constrained mixture model of cardiac growth and remodeling: Analyzing mechanobiological stability and reversal\\
    \hspace{10pt}

    \large
    Amadeus M. Gebauer$^{1,\text{\Letter}}$, Martin R. Pfaller$^{2}$, Fabian A. Braeu$^{3,4}$, Christian J. Cyron$^{5,6}$, Wolfgang A. Wall$^1$ \\
    \hspace{10pt}

    \small
    $^\text{\Letter}$) amadeus.gebauer@tum.de, Tel.: +49 89 289 15255

    $^1$) Institute for Computational Mechanics, Technical University of Munich, 85748 Garching b. München, Germany

    $^2$) Pediatric Cardiology, Cardiovascular Institute, and Institute for Computational and Mathematical Engineering, Stanford University, Stanford, USA\\% falls zu lang reicht auch Pediatrics

    $^3$) Singapore-MIT Alliance for Research \& Technology Center, Singapore

    $^4$) Singapore Eye Research Institute (SERI), Singapore

    $^5$) Institute of Continuum and Materials Mechanics, Hamburg University of Technology, 21073 Hamburg, Germany

    $^6$) Institute of Material Systems Modeling, Helmholtz-Zentrum Hereon, 21502 Geesthacht, Germany

\end{center}

\hspace{10pt}

\normalsize

\section{Details of tissue turnover by constituent}

\begin{figure*}[t]
    \import{supplementary_materials/figures/}{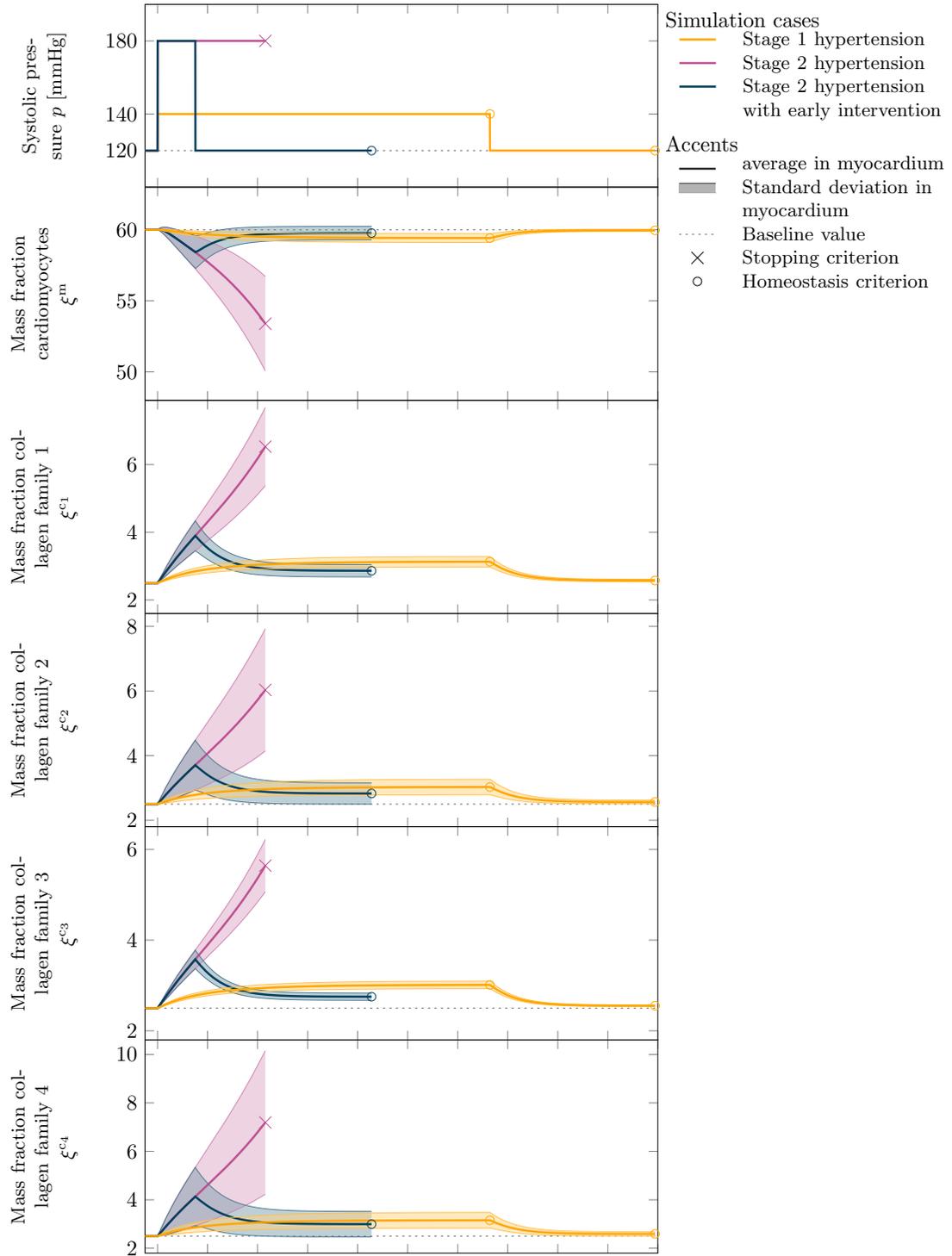}
    \caption{Evolution of the mass fraction of each fiber constituent (cardiomyocytes and collagen fiber families)
    with the standard deviation within the myocardium for the simulation cases
    shown in Figure~\ref{fig:gr_time_curves} of the manuscript.}
    \label{fig:collagen_mass_fraction}
\end{figure*}

Figure~\ref{fig:collagen_mass_fraction} shows the mass fraction of each G\&R fiber (myocytes and 4 collagen fiber families) for the
simulation cases shown in Figure~\ref{fig:gr_time_curves} of the manuscript.
The average mass fraction of each collagen fiber family within the myocardium increase similarly.
Collagen deposition mainly happens near the endocardium. The standard deviation is smaller for collagen
fiber family 3 since this family is lower since the family is approximately oriented in the
longitudinal direction of the heart at the endocardium where most of the collagen is deposited.
The singularity of the fiber direction is, therefore, less pronounced resulting in smaller
deviations within the myocardium.

\begin{figure*}[t]
    \import{supplementary_materials/figures/}{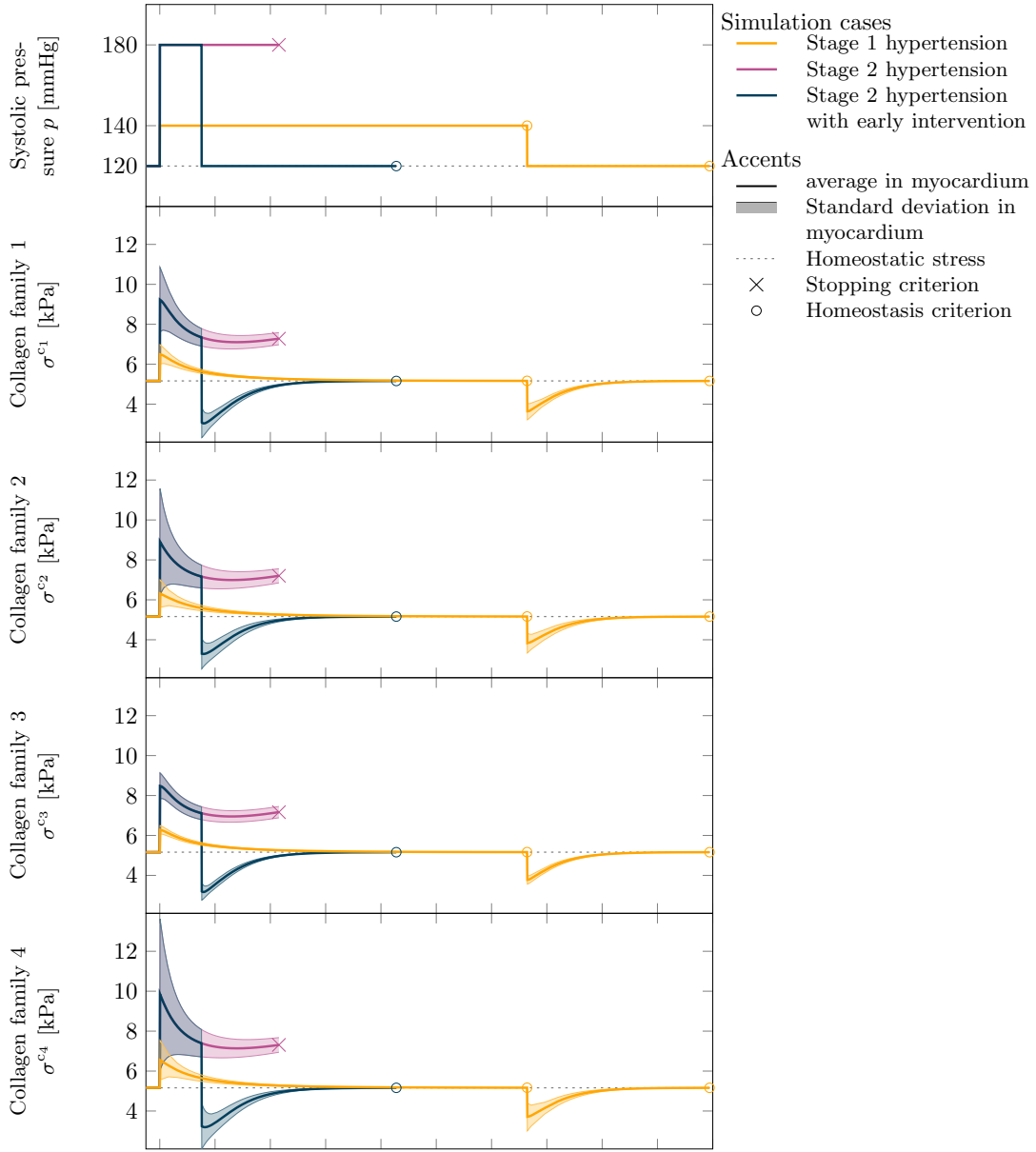}
    \caption{Evolution of the Cauchy stress of all collagen fiber families with the standard deviation within the myocardium for the simulation cases
    shown in Figure~\ref{fig:gr_time_curves} of the manuscript.}
    \label{fig:collagen_stress}
\end{figure*}

Although the mass fraction of cardiomyocytes decreases, the mass itself increases within the myocardium
resulting in an enlargement of individual cardiomyocytes.
Figure~\ref{fig:collagen_stress} shows the fiber stress of each collagen fiber family within the myocardium.
Their convergence behavior is similar to the Cauchy stress of cardiomyocytes reported in the manuscript.
As for the collagen mass fraction, the difference in the standard deviation between the collagen fiber
families stems from the less pronounced effect of the singularity of the fiber orientation at the endocardial apex for collagen fiber family 3.

\clearpage
\section{Influence of spring stiffness}

\begin{figure*}[t]
    \import{supplementary_materials/figures/}{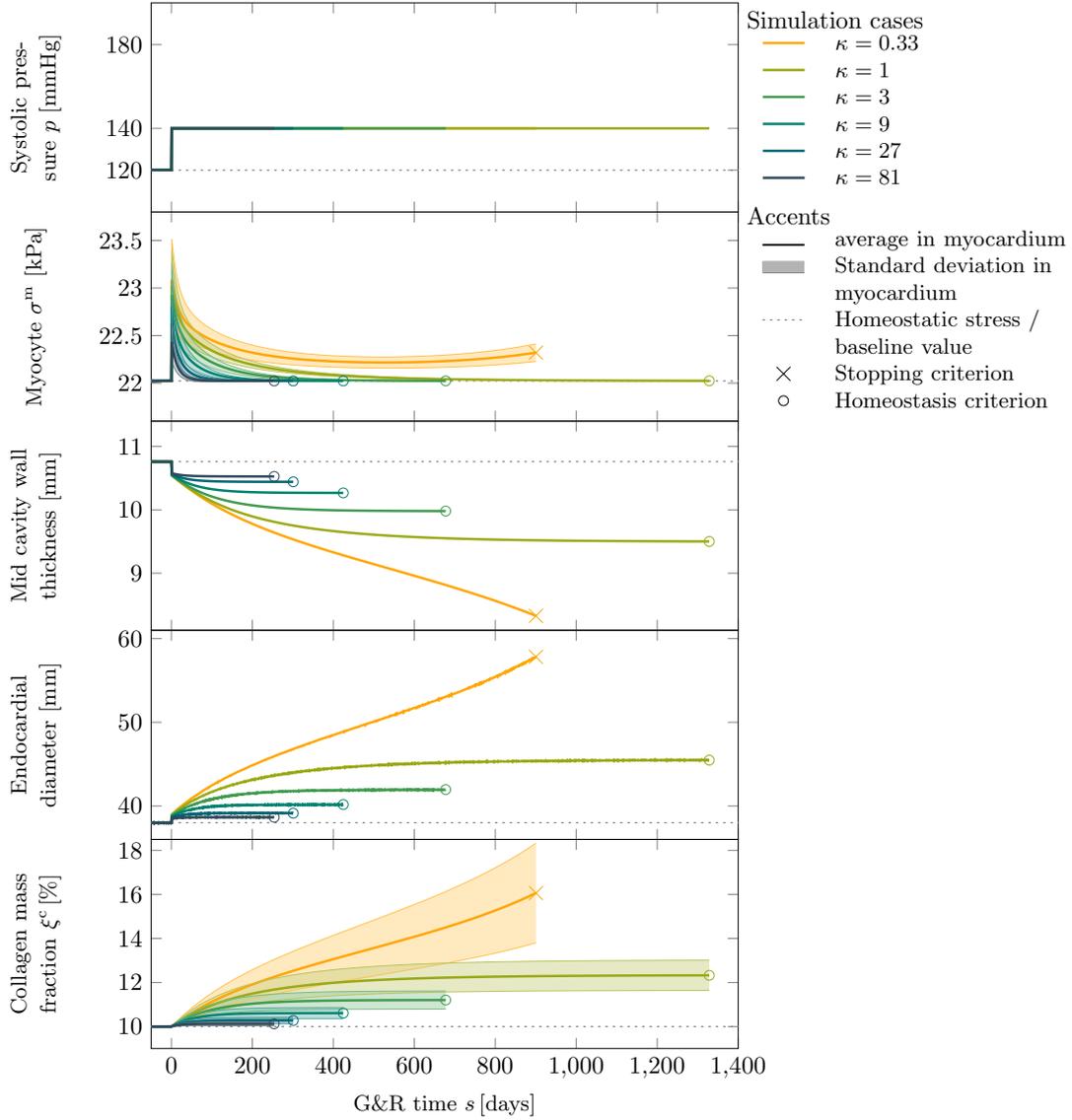}
    \caption{Simulation of stage 1 hypertension with scaled spring stiffness for the boundary condition at the pericardium and the base. Both
    stiffnesses are scaled by $\kappa$. Depicted is the systolic pressure, the myocyte stress with the standard deviation in the myocardium, the mid-cavity
    wall thickness, the endocardial diameter and the collagen mass fraction. For $\kappa=0.33$, the support by the pericardial boundary conditions is
    too small such that G\&R is unstable even for stage 1 hypertension .}
    \label{fig:gr_time_curves_140}
\end{figure*}

The Figures~\ref{fig:gr_time_curves_140} and \ref{fig:gr_time_curves_180} show the influence of the
spring-stiffness on G\&R. The stiffness of the springs at the base and at the epicardium are scaled
by $\kappa$ with $k^iT^i=0.1$. Figure~\ref{fig:gr_time_curves_140} and Figure~\ref{fig:gr_time_curves_180} depict
stage 1 and stage 2 hypertension, respectively.
The deviation from homeostasis gets smaller for stiffer spring boundary conditions. As a result,
G\&R is less pronounced (smaller endocardial diameter and thinner mid-cavity wall thickness).
This also influences mechanobiological stability. For $k=0.33$, even (previously stable) stage 1 hypertension results
in unstable G\&R as the pericardium provides too little support for stabilization. In contrast, (previously unstable) stage 2 hypertension
can result in stable G\&R for $\kappa \ge 9$.

\begin{figure*}[t]
    \import{supplementary_materials/figures/}{p180_peri_time_curves.tex}
    \caption{Simulation of stage 2 hypertension with scaled spring stiffness for the boundary condition at the pericardium and the base. Both
    stiffnesses are scaled by $\kappa$. Depicted is the systolic pressure, the myocyte stress with the standard deviation in the myocardium, the mid-cavity
    wall thickness, the endocardial diameter and the collagen mass fraction. For a large enough stiffness ($\kappa \ge 9$), stage 2 hypertension can also
    result in stable G\&R.}
    \label{fig:gr_time_curves_180}
\end{figure*}